\documentclass[twocolumn,superscriptaddress,floatfix,showpacs]{revtex4-1}
\pdfoutput=1
\usepackage{graphics,amssymb,amsmath,epsfig,color}
\usepackage{graphicx}

\newcommand{\be}{\begin{equation}} \newcommand{\ee}{\end{equation}}
\newcommand{\bea}{\begin{eqnarray}} \newcommand{\eea}{\end{eqnarray}}
\begin{document}

\title{Two-dimensional SIR epidemics with long range infection}
\author{Peter Grassberger} \affiliation{JSC, FZ J\"ulich, D-52425 J\"ulich, Germany}
         \affiliation{Max Planck Institute for the Physics of Complex Systems, N\"othnitzer Strasse 38, D-01187 Dresden, Germany}
\date{\today}

\begin{abstract}

We extend a recent study of susceptible-infected-removed epidemic processes with long range 
infection (referred to as I in the following) from 1-dimensional lattices
to lattices in two dimensions. As in I we use hashing to simulate very large 
lattices for which finite size effects can be neglected, in spite of the assumed 
power law $p({\bf x})\sim |{\bf x}|^{-\sigma-2}$ for the probability that a site can infect
another site a distance vector ${\bf x}$ apart. As in I we present detailed results 
for the critical case, for the supercritical case with $\sigma = 2$, and for the supercritical
case with $0< \sigma < 2$. For the latter we verify the stretched exponential growth of the 
infected cluster with time predicted by M. Biskup. For $\sigma=2$ we find generic power laws
with $\sigma-$dependent exponents in the supercritical phase, but no Kosterlitz-Thouless (KT)
like critical point as in 1-d. Instead of diverging exponentially with the distance from the 
critical point, the correlation length increases with an inverse power, as in an ordinary 
critical point. Finally we study the dependence of the critical exponents on 
$\sigma$ in the regime $0<\sigma <2$, and compare with field theoretic predictions. 
In particular we discuss in detail whether the critical behavior 
for $\sigma$ slightly less than 2 is in the short range universality class, as conjectured
recently by F. Linder {\it et al.}. As in I we also consider a modified version of the 
model where only some of the contacts are long range, the others being between nearest 
neighbors. If the number of the latter reaches the percolation threshold, the 
critical behavior is changed but the supercritical behavior stays qualitatively the same.

\end{abstract}

\maketitle

\section{Introduction}

Epidemic spreading processes have attracted 
increasing attention in the statistical physics community \cite{Newman02}. 
In the simplest case of short range infection, no cooperative effects \cite{Janssen04,Bizhani},
and `removal' after the infection (i.e. either immunization or death), this `general epidemic
process' \cite{Mollison,Grass83} generates ordinary percolation clusters. In the present 
paper we shall deal with the generalization to the case where the process lives on a 2-d 
square lattice and at least some of the infections have long range. More precisely we 
shall study cases where each site can infect a finite number of other sites, and the 
probability for an infection to `jump' a distance vector ${\bf x}$ decreases for large 
${\bf x}$ as $p({\bf x})\sim |{\bf x}|^{-\sigma-d}$ with $d=2$.

Models of this type were first suggested by \cite{Mollison}. A first calculation of 
critical exponents in \cite{Grassberger86} was flawed, as pointed out by Janssen \cite{Janssen99},
and numerical verifications of the exponents calculated in \cite{Janssen99} where 
presented in \cite{Linder}. 

These works were done in the spirit of the renormalization group (RG) for critical phenomena,
and have to be seen in the context of other models with long range interactions, most 
prominently the Ising model \cite{Fisher72,Sak,Yamazaki,Gusmao,Honkonen,Luijten01,Luijten02}.
As we shall discuss later in detail, they give most unambiguous results for $\sigma \approx 0$,
where analytic results can be obtained by means of perturbative field theoretic RG. 

Related to this was work on the case $\sigma=d$, which started already very early 
with papers by Ruelle \cite{Ruelle}, Dyson \cite{Dyson}, Anderson \cite{Anderson}, 
Thouless \cite{Thouless} and others. The main result obtained there was that the 1-dimensional 
Ising model with $\sigma=1$ not only has a phase transition, but that the critical point
is of Kosterlitz-Thouless (KT) type: Below the critical temperature one 
finds power laws with continuously varying exponents, while near the critical point 
the correlation length increases exponentially with the inverse distance from it. 
The most detailed numerical verification of the Ising model predictions is in \cite{Luijten01}.
Analogous mathematical results for the Potts model and for percolation were obtained in 
\cite{Cardy,Schulman,Aizenman,Aizenman88}, but where not tested numerically for more than two 
decades, because of the obvious problem of simulating the very large lattices required
to overcome finite size effects. They were verified only recently in \cite{Grass13}
(denoted as I in the following), where hashing was used to simulate 1-d lattices of 
size $L=2^{64}$. 

Independently of the above mentioned work on SIR epidemics with long range 
contacts leading to power laws, increasingly much interest has been devoted recently 
to {\it supercritical} epidemics with $0<\sigma<d$. Partly this comes from the interest 
in the navigability of small world networks \cite{Kleinberg}, partly from interest 
in the spreading of various agents (viruses like influenza or HIV, computer viruses,
rumors, and even money \cite{Brockmann06,Brockmann}) on real world networks. But even
apart from such epidemic-like processes, spatial networks with long range links 
have become powerful paradigms for complex real world systems \cite{Barthelemy}.
Supercritical SIR epidemics with long range infection provide the easiest and 
most straightforward way to generate them.

The main objective of the present paper is to extend the simulations of I to 
two spatial dimensions. Again we shall use hashing, in order to simulate 
square lattices with $2^{64}$ sites. We could have easily increased this to 
$2^{128}$ sites with only minor loss of efficiency, but we checked that finite 
size effects can be safely neglected already with $2^{64}$ sites, for nearly all values 
of $\sigma$ (except for $\sigma$ very close and below zero) and for all observables. 
While these simulations have no finite {\it lattice} size corrections, there are of 
course finite {\it cluster} size corrections, or eqivalently corrections to scaling 
due to the finite duration of the epidemic. We thus complement these simulations also
with simulations on finite lattices (up to $65536\times 65536$) where we followed 
all epidemics until they died.
In Sec.~2 we shall define the models in detail, and sketch the main results. Supercritical
epidemics with $0<\sigma<2$ will be discussed in Sec.~3, while the case $\sigma=2$
is treated in Sec.~4. Finally, the critical case is studied in Sec.~5, and 
our conclusions are drawn in Sec.~6.

\section{The model and basic features}

We shall only deal with basic SIR epidemics on the square lattice with periodic
boundary conditions. In all simulations the lattice size is $2^{64}$ sites, time is 
discrete and the infective period is one time step, after which sites become immune. 
We start always with a single infected site at $t=0$, all other sites being 
susceptible. In each time step every infected site tries to infect in 
average $k_{\rm out}$ other sites. If these sites are no longer susceptible, no
replacements for them are chosen, i.e. the number of new infections is just reduced.

As in I, we consider two different models as how the target sites for new infections are 
selected. Most of the simulations were done for model (A), where each target site with
distance vector ${\bf x}$ from the infectious site is chosen randomly with probability
$p({\bf x})$. More precisely, we select first $k_0 =\lfloor k_{\rm out} \rfloor$ such 
sites, and then one more site is chosen with probability $p=k_{\rm out}-k_0$.

In model (B) (studied only in Sec.~IV) we first infect each of the four neighbors with 
probability $q\in [0,1]$, and then choose $k_{\rm out}-4q$ additional targets as in model 
(A). For $q>1/2$ this means that the process is always supercritical, and for $q=1$ it
implies that every epidemic always leads to an infinite cluster. We include this 
model for comparison with results in the mathematical literature 
\cite{Biskup-2004,Benjamini-2001,Berger-2004,Coppersmith}, where the emphasis was 
on supercritical long range percolation.

Following Linder {\it et al.} \cite{Linder}, we define $p({\bf x})$ for $\sigma >0$ implicitly 
by the following simple algorithm:\\
(i) We first chose two random numbers $u,v$ uniformly from $]0,1]\times ]0,1]$.\\
(ii) If $w^2 \equiv u^2+v^2 >=1$, we discard them and choose a new pair until $w^2 <1$.\\
(iii) Finally, we define 
$$ x=\pm\;\frac{u}{w^{1+2/\sigma}},\quad y=\pm\;\frac{v}{w^{1+2/\sigma}} $$
where all four sign combinations are chosen with equal probability. 
(iv) When computing the position of the new infected site, we added {\bf x} and used 
periodic boundary conditions.

For $\sigma <=0$ this can be modified suitably, but we shall not give any 
details as we will in the following show only simulations for $\sigma >0$. 
These simulations took in total about four CPU years on modern PCs.

\begin{figure}
\begin{center}
\includegraphics[width=0.57\textwidth, bb=50 0 880 550]{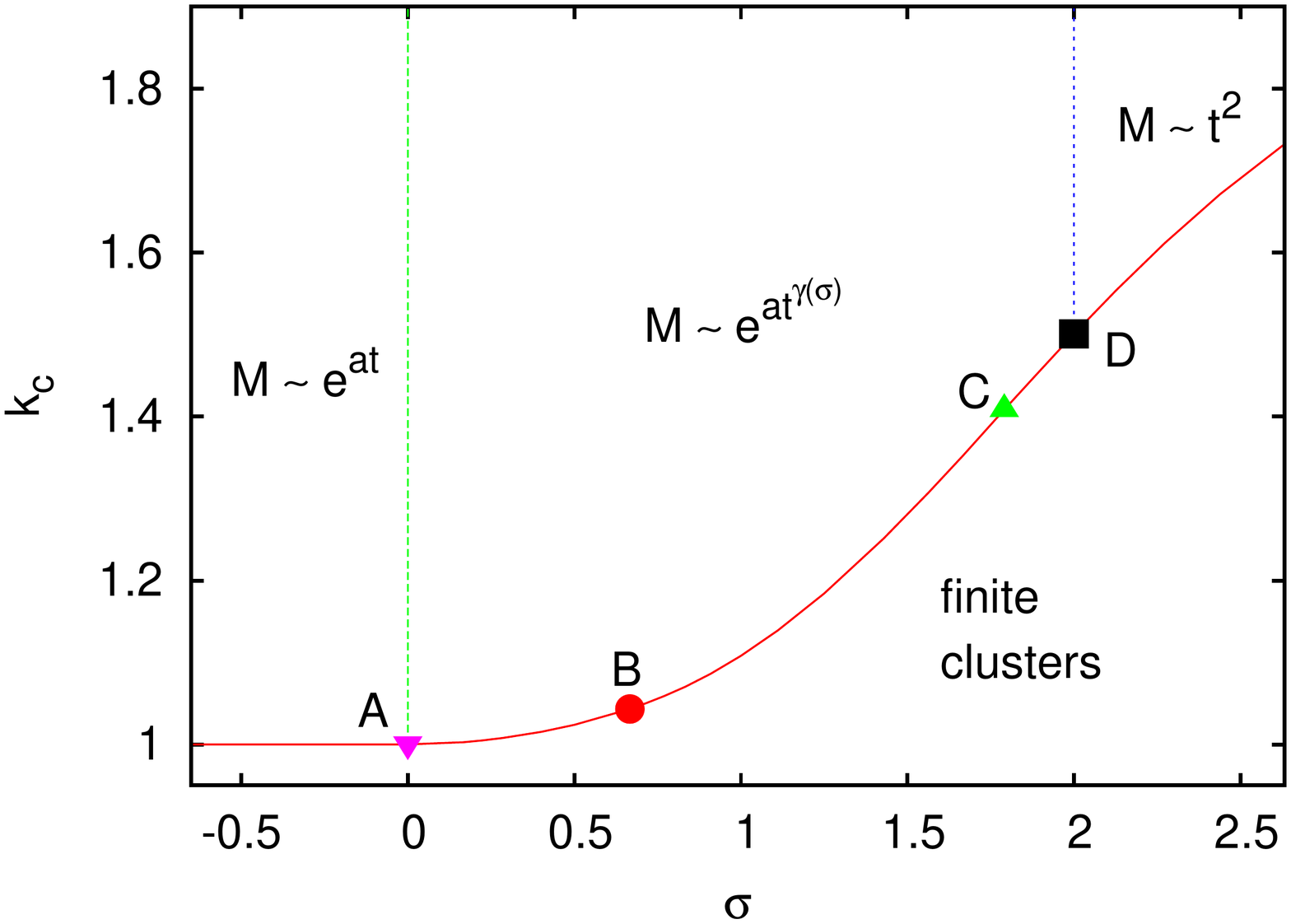}
\caption{(Color online) Phase diagram for model (A). Below the continuous curve, i.e. 
for small values of $k_{\rm out}$, there are only finite clusters with probability 1.
Infinite epidemics can exist only above this critical curve $k_{\rm out} = k_c(\sigma)$. 
For supercritical epidemics we have three regimes: For $\sigma <0$ (i.e., to the left 
of the left dashed line) the process is of {\it mean field} type; for $0<\sigma <2$ 
(between the dashed lines) it is of {\it intermediate} type; and for $\sigma >2$ it is 
basically as in short range percolation. For critical epidemics, one has three 
analogous regimes, but the boundaries between them are different. The boundary between 
mean field and intermediate-range epidemics is at point B, while the one between 
intermediate-range and short range is either at point D or near point C.}
   \label{Phases}
\end{center}
\end{figure}

The phase diagram for model (A) is shown in Fig.~\ref{Phases}.
Below the continuous curve, i.e.
for small values of $k_{\rm out}$, there are only finite clusters with probability 1.
Infinite epidemics can exist only above this critical curve $k_{\rm out} = k_c(\sigma)$. 
For $\sigma <0$ clusters are tree-like, i.e. the probability that a given site is one 
a finite loop decreases as an inverse power of the lattice size $L$. This implies 
in particular $k_c =1$ for $\sigma <0$ (for $\sigma$ slightly larger than 0, we found 
numerically that the critical curve scales as $k_c-1 \sim\sigma^2$).
The masses of supercritical clusters with $\sigma <0$ (i.e., to the left of the left
dashed line) increase thus exponentially with time. In contrast, for $\sigma >2$ (to
the right of the the right hand dashed line) supercritical clusters are essentially
compact with linearly increasing size, i.e. their masses increase $\sim t^2$. In
between, for $0<\sigma<2$, the masses of supercritical clusters increase as stretched
exponentials with $\sigma-$dependent exponents, as proven in \cite{Biskup-2004,Biskup-2009}. 
In general, for $d$ dimensions of space, the boundaries between mean field, intermediate
range and short range are at $\sigma=0$ and $\sigma=d$ \cite{Biskup-2004,Biskup-2009}.

For {\it critical} epidemics, however, the  
boundaries between these {\it mean field}, {\it intermediate}, and {\it short range} regimes 
are different. Critical exponents are mean field like for all $\sigma < 2/3$ ($\sigma < d/3$
for $d$ spatial dimensions), i.e. to 
the left of point B \cite{Janssen99,Linder}. For $\sigma >2/3$ all critical exponents 
depend continuously on $\sigma$, until the {\it short range} regime is reached. 
According to \cite{Linder}, this happens at point C which corresponds to 
$\sigma_C = 43/24 = 1.79166\ldots$ ($\sigma_C = d-2\beta/\nu$ in general), where 
$\beta$ and $\nu$ are the order parameter and correlation length critical exponents 
for ordinary (short-range) percolation). 
Our own simulations, presented in Sec. 5, suggest 
that the short range regime extends only down to $\sigma=2$. The detailed critical
behavior for $\sigma_C < \sigma \leq 2$ is unclear.

\begin{figure}
\includegraphics[width=0.53\textwidth]{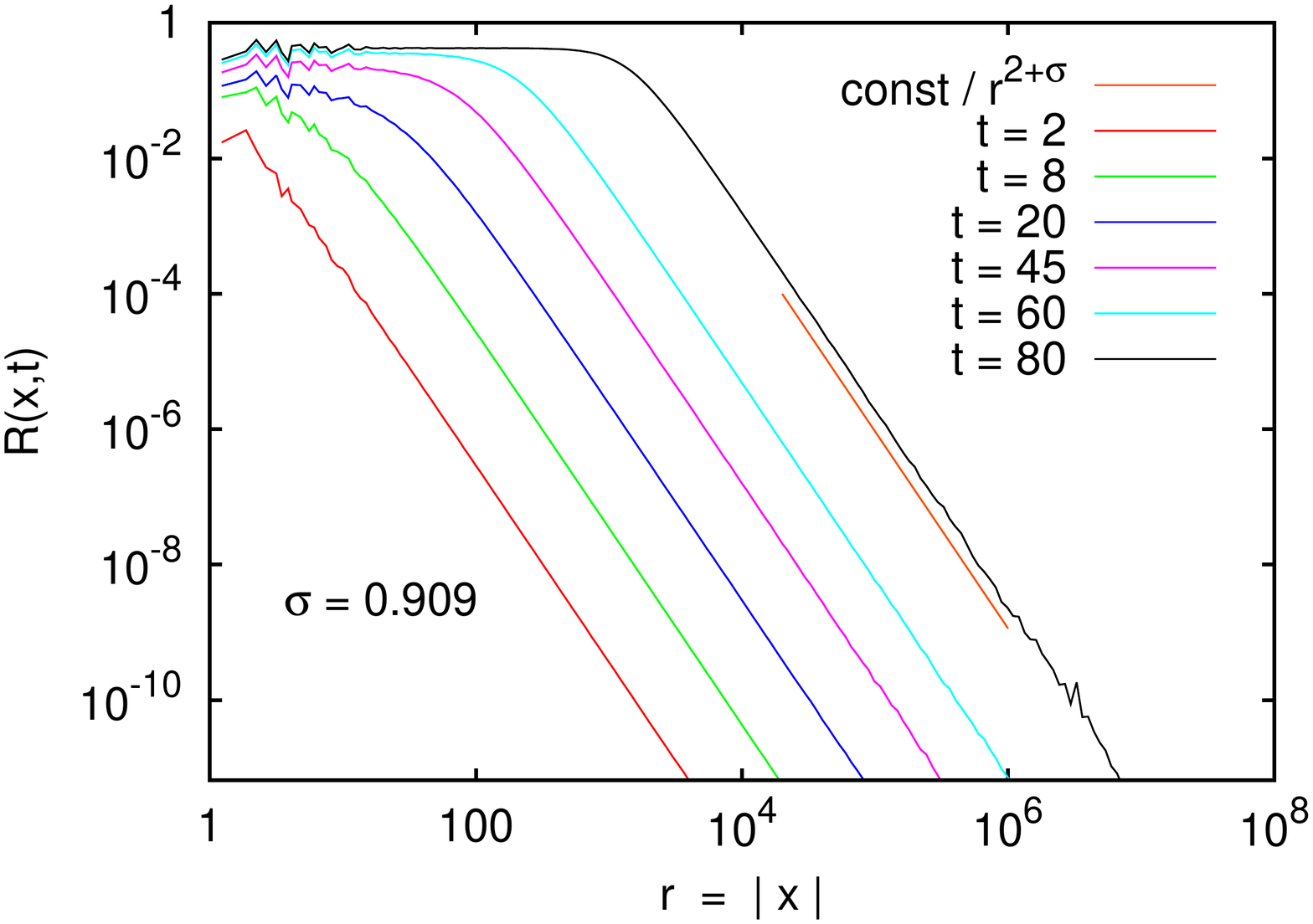}
\caption{(Color online) Densities of removed sites for model (A) with $\sigma = 0.909$
and $k_{\rm out} = 1.4$, at four values of $t$. The fluctuations at small $r$ are lattice 
artifacts and should not be taken serious, as also in the next figure and in Fig.~\ref{front-1.3}.}
\label{front-1.60}
\end{figure}

\begin{figure}
\includegraphics[width=0.53\textwidth]{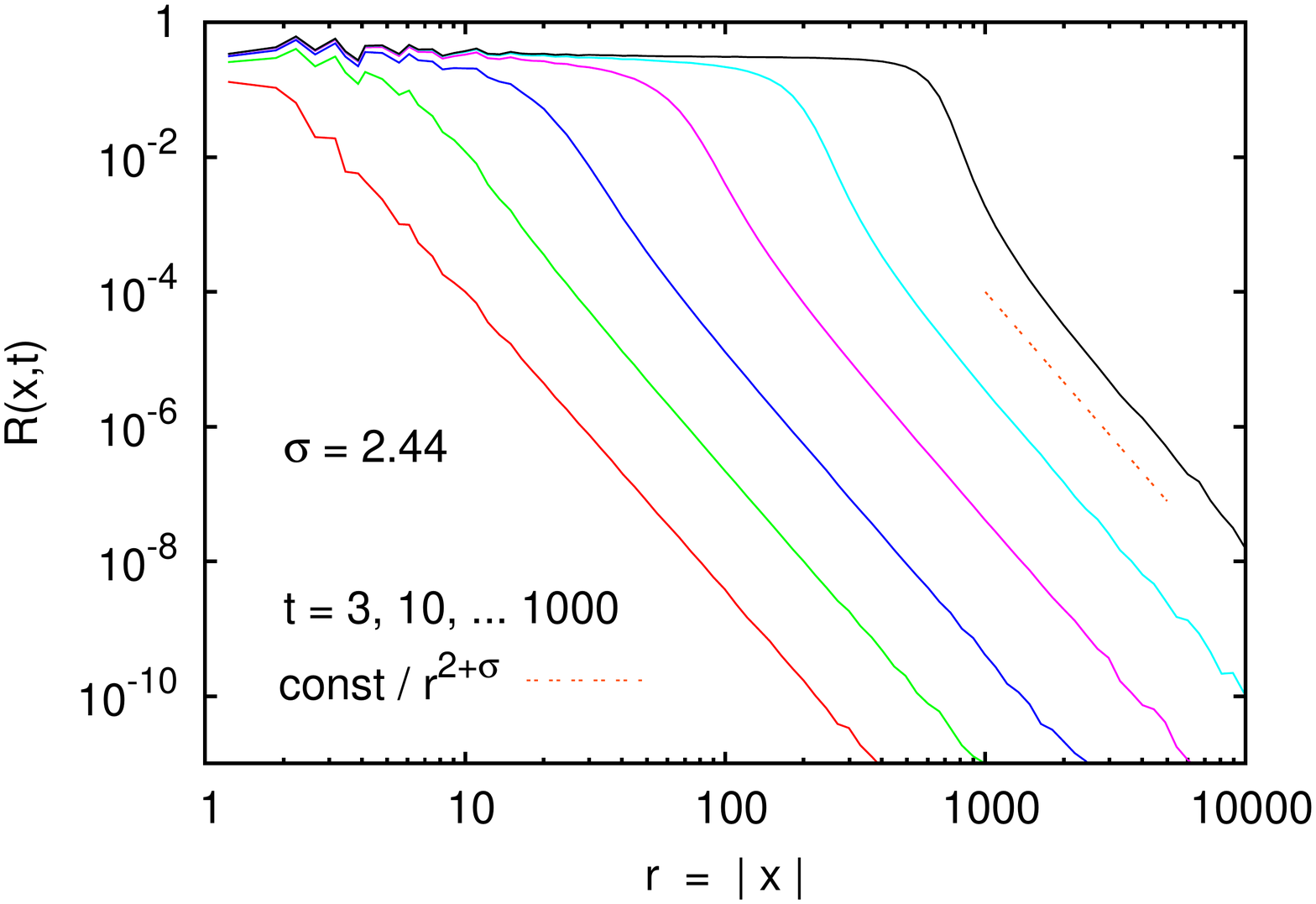}
\caption{(Color online) Densities of removed sites for model (A) with $\sigma = 2.44$
and $k_{\rm out} = 1.7$, at six values of $t$. Notice that in this case clusters 
have increasingly sharp edges on a logarithmic scale, followed by a increasingly weak
tail where the density decays according to Eq.~(\ref{tail}).}
\label{front-0.91}
\end{figure}

At $\sigma=0$ and $\sigma=d$ also the spatial structure of clusters of removed sites changes
qualitatively. Assume that the epidemic started at ${\bf x}=0$. Then the density of 
removed sites $R({\bf x},t)$ decreases for any $t>0$ asymptotically as 
\be
   R({\bf x},t) \sim |{\bf x}|^{-\sigma-d}  \quad {\rm for} \;\;\; |{\bf x}| \to\infty
             \label{tail}
\ee
for all $\sigma >0$, while $R({\bf x},t)$ becomes increasingly flat when $\sigma < 0$. 
The latter is illustrated for $d=1$ in Fig.~1 of I, while the former is illustrated 
in Fig.~\ref{front-1.60}.

Eq.~(\ref{tail}) holds also for $0<\sigma>2$ when $|{\bf x}| \gg t$, but at smaller 
distances the density decays faster than a power law, see Fig.~\ref{front-0.91}. At
least for supercritical epidemics this transition between the behaviors illustrated
in Figs.~\ref{front-1.60} and \ref{front-0.91} happens exactly at $\sigma=2$. Whether
this is still true for critical epidemics will be discussed in Sec. 5.

\section{Supercritical Behavior for $0 < \sigma < 2$.}

Although most studies of percolation and epidemic processes in the physics literature deal 
only with critical cases, it is clear that supercritical processes are of utmost practical
importance. Indeed, it can be argued that most real epidemics like rabies, HIV, or 
various strands of influenza are far supercritical, take place in two dimensions of 
space, and are transmitted -- due to long distance travels -- by effective contacts with 
long range. Accordingly, there are several papers in the recent physics literature 
where either the spreading or the topology of the generated clusters were studied. 

Typical examples for the former are \cite{Mancinelli,Castillo} and \cite{Brockmann}, 
where front propagation was studied by mean field type arguments. While a finite
speed of propagation was claimed in \cite{Brockmann}, velocities that increase 
exponentially with time were found in \cite{Mancinelli,Castillo}. As regards the 
topology of the cluster of removed sites, the main discussion was whether it forms 
a small world network (with graph diameter increasing logarithmically with the 
number of sites), has dimension two (as for short range epidemics), or is in between 
\cite{Sen01,Sen02,Moukarzel,Goswami11,Emmerich}.

As seen from Fig.~\ref{front-1.60}, the spatial structure of cluster is rather 
simple. For small distances from the seed, there is a region of constant density
smaller than 1, while there is a power law decay at larger densities. Indeed, 
Fig.~\ref{front-1.60} and similar plots for other values of $\sigma \in [0,2]$
are compatible, for large $t$, with a scaling form 
\be
   R({\bf x},t) \sim \phi(|{\bf x}|/\xi(t)) 
             \label{super-dens}
\ee
with 
\be
   \phi(z) = \left\{
            \begin{array}{rl}
            \text{ const } & \text{ for } z \ll 1,\\
            z^{-2-\sigma}  & \text{ for } z \gg 1 ,
            \end{array} \right.
     \label{super-dens1}
\ee
and a smooth cross-over between the two regimes. The fact that $\phi(z) <1$ for 
all $z$ can be shown easily. Indeed, since the chance for site ${\bf x}$ to get 
infected by site ${\bf y}$ is $p({\bf y-x})$, the probability for it to have 
never been infected is 
\be
   1-\lim_{t\to\infty} R({\bf x}) \geq \prod_{\bf y}(1-p({\bf y-x})),
\ee
giving
\be 
   \log [1-\lim_{t\to\infty} R({\bf x})] \geq \sum_{\bf y} \log[1-p({\bf y})] > -\infty.
\ee
According to Eq.~\ref{super-dens}, the function $\xi(t)$ also controls the geometric
average radial size, 
\be
   r(t) \equiv \langle |{\bf x}(t)|\rangle_{\rm geom} \equiv \exp(\langle \ln |{\bf x}(t)|\rangle \sim \xi(t).
\ee
It also controls the size of the cluster of infected sites at any given time, which is a fuzzy 
ring of radius $\xi(t)$. 

If the size would increase exponentially with time, as obtained in mean field 
theory \cite{Mancinelli,Castillo}, this would mean that the cluster has the small 
world property, since its graph diameter is $\leq 2t$. On the other hand, if the 
radius would increase linearly \cite{Brockmann}, its dimension would be $\leq 2$.
Actually, as proven rigorously in \cite{Biskup-2004,Biskup-2009}, neither is correct.
Instead, the size increases for any dimension $d$ like a stretched exponential,
\be
   r(t) \sim \exp(at^\gamma)     \label{stretch}
\ee
with 
\be
   \gamma = \gamma(\sigma) = \frac{1}{\log_2 \frac{d+\sigma}{2d}}. \label{gamma}
\ee
For $\sigma \to 0$ one has $\gamma \to 1$, i.e. one obtains the small world behavior 
of mean field theory. For $\sigma \to 2$, on the other hand, $\gamma \to 0$. 
Qualitatively similar behavior was suggested in \cite{Emmerich} on the basis of 
numerical simulations, but the detailed functional form of $\gamma(\sigma)$ 
obtained in \cite{Emmerich} was different. 

\begin{figure}
\includegraphics[width=0.57\textwidth]{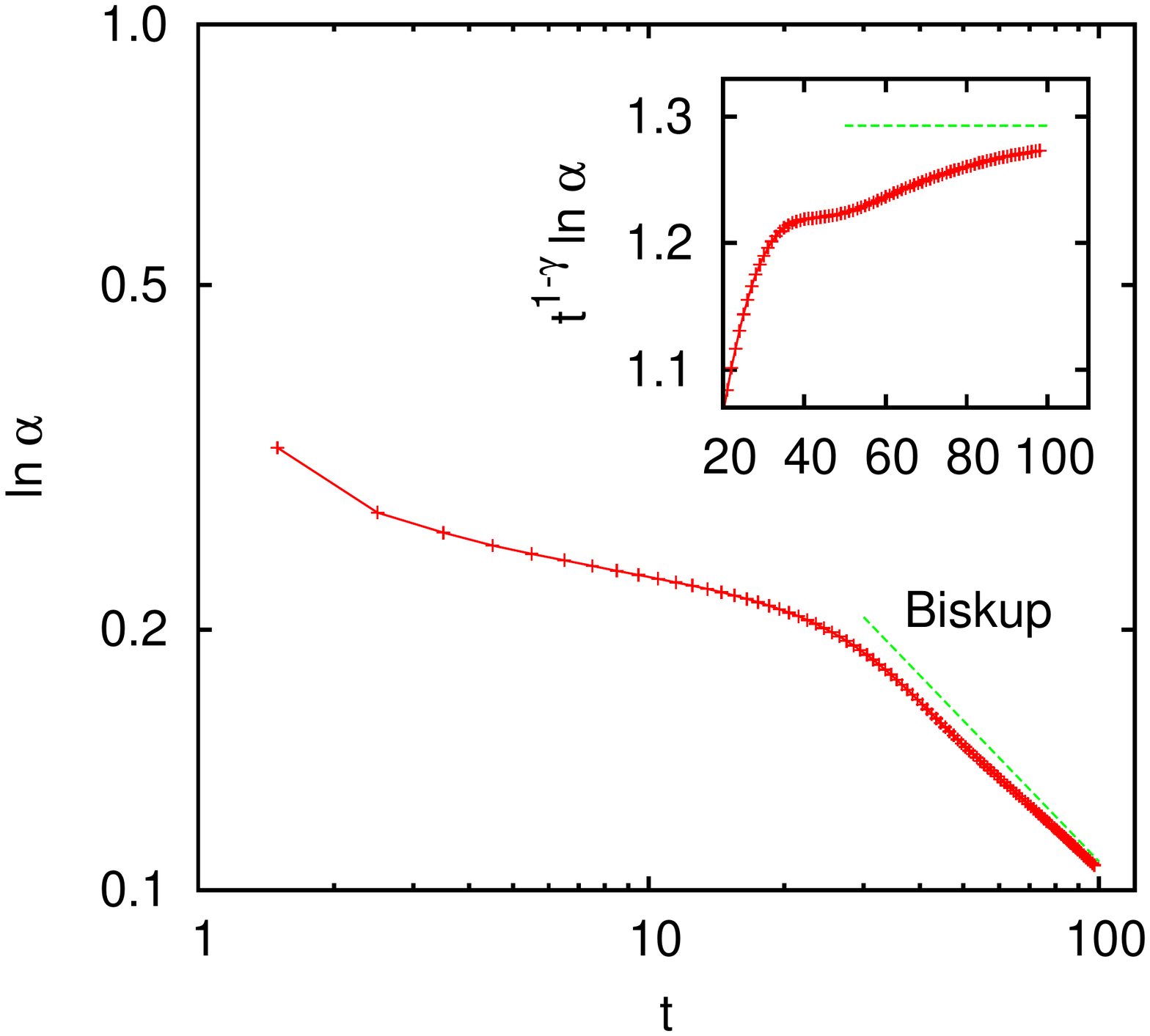}
\caption{(Color online) Log-log plot of the effective growth rate $\alpha(t)$ for 
$\sigma = 0.909$ and $k_{\rm out} = 1.4$. The dashed straight line is the prediction 
of \cite{Biskup-2004}. Statistical errors are smaller than the line width. The insert
shows the same data (on a linear plot) after multiplication with $t^{1-\gamma}$.}
\label{gamma-1.6}
\end{figure}

Verifying Eqs.~(\ref{stretch}) and (\ref{gamma}) numerically is not easy: 

\begin{itemize}

\item First of all,
for finite $t$ slightly different results are obtained when $r(t)$, the number $n(t)$
of active sites, or the number $N(t)$ of removed (`immune') sites is considered, 
although they all should show the same asymptotic behavior up to powers of $t$. In 
the following we shall concentrate on $n(t)$. 

\item Secondly, as found also in I, directly 
fitting $n(t)$ with a stretched exponential gives much too large estimates for 
$\gamma$. It is much better to define an effective growth rate 
\be
   \alpha(t) = \ln \left[\frac{n(t+1/2)}{n(t-1/2)}\right].     \label{gr-rate}
\ee
which should, according to Eq.~(\ref{stretch}), decrease as $\alpha(t) \sim t^{\gamma-1}$.

\item For small values of $\sigma$,
$n(t)$ does increase exponentially with $t$ for very long times, if $k_{\rm out}$ is 
not very large. The reason is that the deviation from exponential increase is due to 
saturation effects, and these set in very late for small $\sigma$ and $k_{\rm out}$.
This is clearly seen in Fig.~\ref{gamma-1.6}, where $\alpha(t)$ is plotted against $t$
on a log-log plot. The straight dashed line is the prediction of 
Eqs.~(\ref{stretch}) and (\ref{gamma}), and it gives a decent fit only for $t>30$.
Due to this effect, we were not able to verify Eq.~(\ref{gamma}) for $\sigma <0.3$.

\item As seen in the insert in Fig.~\ref{gamma-1.6}, even for large values of $t$ 
there are strong systematic deviations from the predicted asymptotic behavior 
(notice that statistical errors in Fig.~\ref{gamma-1.6} are much smaller that the 
line width). Similar systematic corrections were also found for all other values of 
$\sigma$ and $k_{\rm out}$. It seems that they were underestimated in I and are 
responsible for most of the systematic overestimation of $\gamma$ found in that 
paper.
\end{itemize}

\begin{figure}
\includegraphics[width=0.5\textwidth]{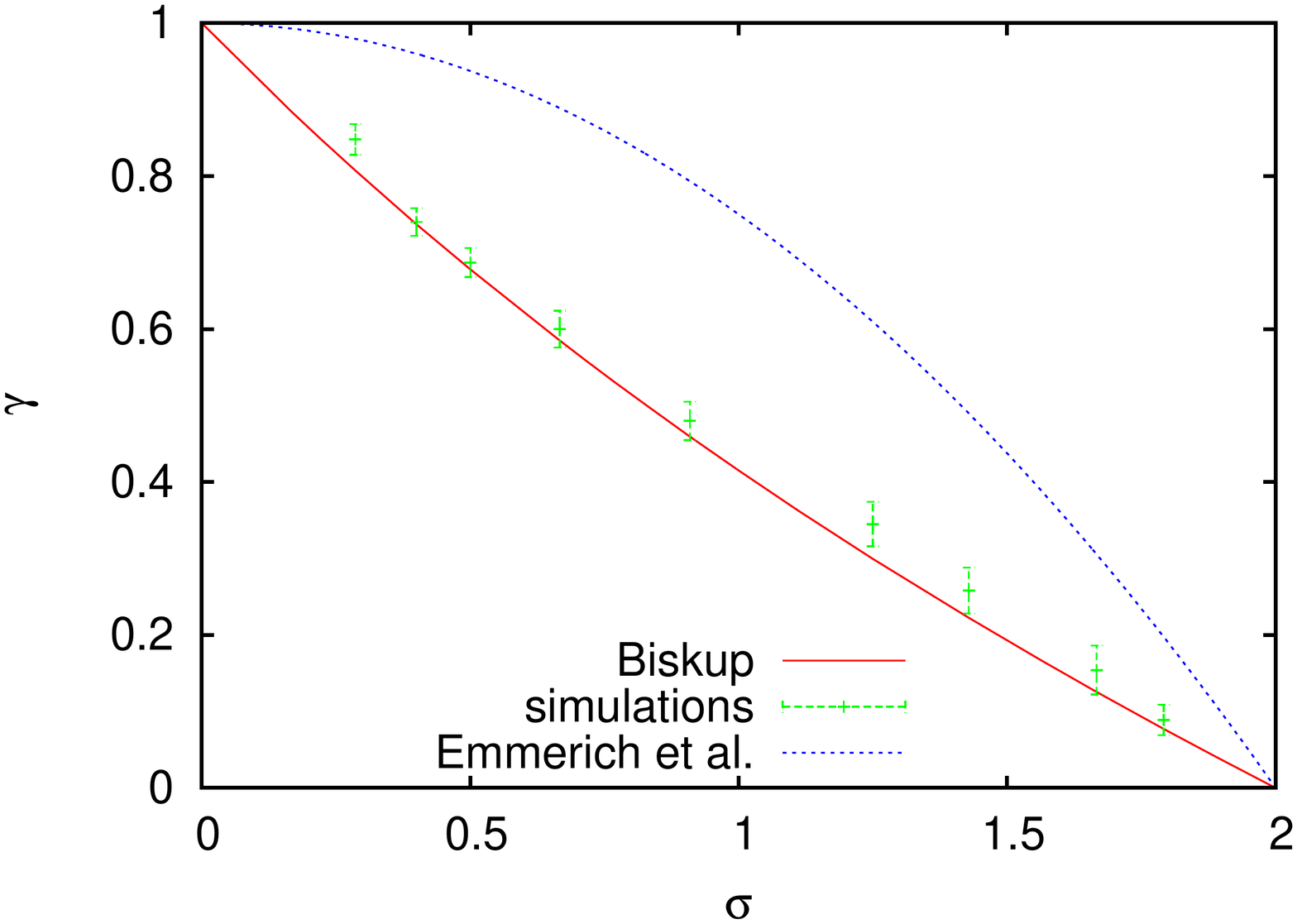}
\caption{(Color online) Measured values of $\gamma$ versus $\sigma$, compared to
the predictions of Biskup \cite{Biskup-2004,Biskup-2009} and Emmerich et al. 
\cite{Emmerich}. Error bars represent highly subjective estimates of systematic
errors, while purely statistical errors would be negligible.}
\label{stretch.fig}
\end{figure}

In spite of these problems, our final results shown in Fig.~\ref{stretch.fig}
are fully compatible with Eqs.~(\ref{stretch}), (\ref{gamma}) and definitely rule out the 
alternative conjecture of \cite{Emmerich}.

\section{The case $\sigma = 2$}

In the supercritical phase, the transition between intermediate and short-range 
behaviors happens at $\sigma = d$. For $d=1$ we found in I that the number of infected
sites increases for $\sigma = 1$ with a power law 
\be
   n(t,k_{\rm out}) \sim t^{\eta(k_{\rm out})},     \label{super}
\ee
provided $k_{\rm out}$ was larger than a critical value $k_c$. The power $\eta$
depended continuously on $k_{\rm out}$, and the behavior near the critical point $k_c$
was of Kosterlitz-Thouless type with the correlation length (and time) increasing
like a stretched exponent with the distance from criticality. The order parameter 
(the density of the infinite cluster) was a constant $>0$ at criticality, showing 
that the transition is first order.

Indeed, except for details this behavior for $d=1$ had been predicted long ago, 
and some aspects had even been proven rigorously \cite{Cardy,Schulman,Aizenman,Aizenman88}.
Comparably detailed predictions are not available for $d=2$. It was conjectured 
in \cite{Benjamini-2001} that scaling with continuously varying exponents as in 
Eq.~(\ref{super}) holds for model (B) with $q=1$ (i.e., when every site is connected
to each of its neighbors, so that $k_c=4$), but neither the dependence of 
$\eta$ on $k_{\rm out}$ nor the behavior in model (A) are known.

\begin{figure}
\includegraphics[width=0.53\textwidth]{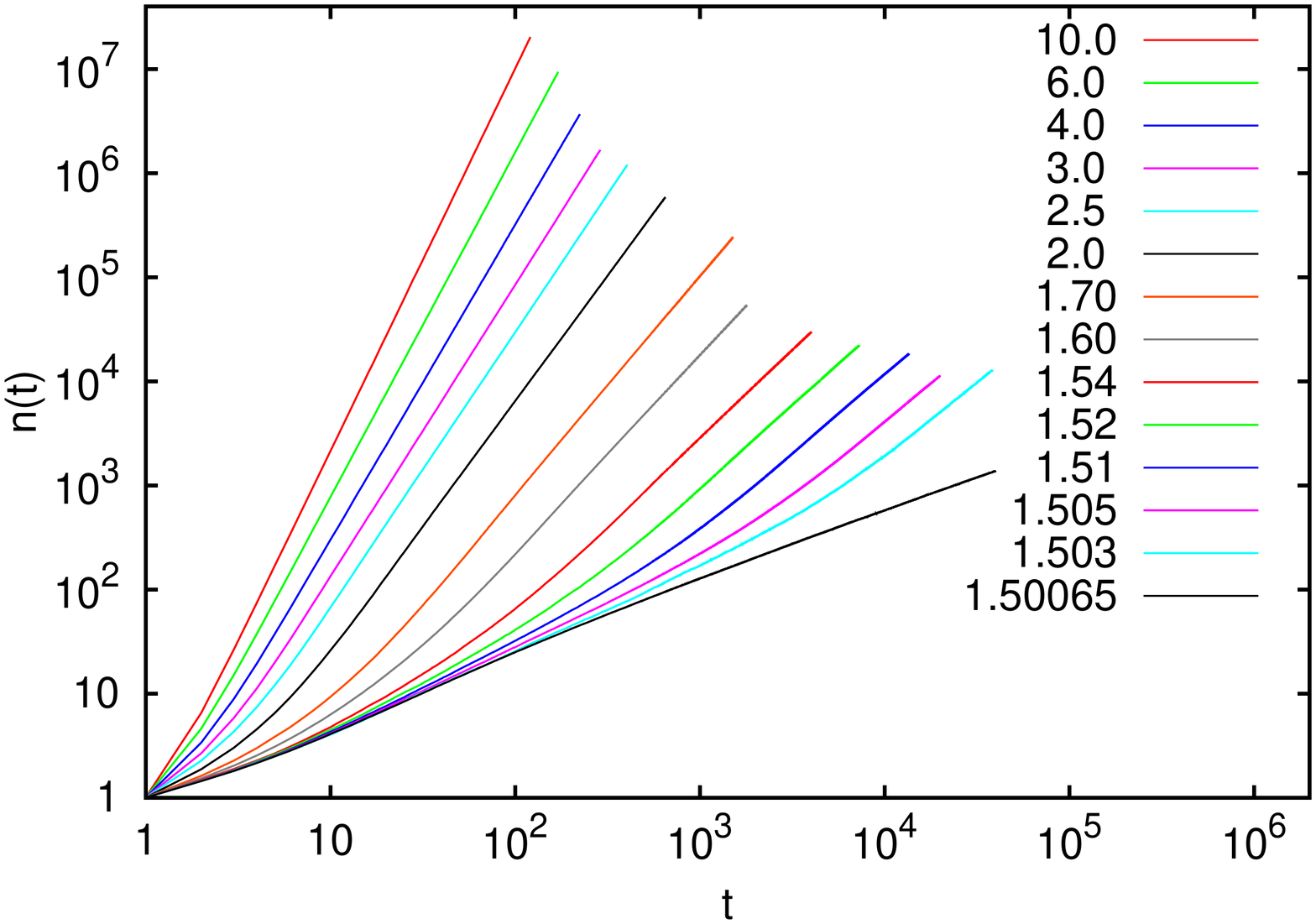}
\caption{(Color online) Log-log plots of $n(t)$ versus $t$ for several values of 
$k_{\rm out}$. These values are given on the r.h.s. of the figure, with the uppermost
curve corresponding to the largest $k_{\rm out}$. Statistical errors are comparable to 
or smaller than the line width.}
\label{n-t-sigma2}
\end{figure}

Values of $n(t)$ for model (A) and for different values of $k_{\rm out}$ are shown in
Fig.~\ref{n-t-sigma2}. We see that indeed all curves for $k_{\rm out}\geq 1.503$ show
power laws for large $t$, with exponents decreasing with $k_{\rm out}$. As $k_{\rm out}$
approaches $k_c\approx 1.5007$ from above, this asymptotic power law sets in later 
and later. At the same time, in this limit a different power law is observed for small $t$,
with an exponent $\eta(k_c) \approx 0.6$ which is compatible within errors with the 
exponent for critical SIR epidemics with short range contacts \cite{Grass-1992} (a 
more detailed comparison with short range SIR epidemics will be given in the next section). 

For model (B) with $q > 0.5$ the behavior is slightly different. In that case the 
process is always supercritical, and  thus the asymptotic power laws hold down to $t = O(1)$.
There is no time regime for small $k_{\rm out}$ where the asymptotic power law is replaced 
by a different power law. This is not true for model (B) with $q =0.5$. Although the process
is also in this case supercritical for any number of long range contacts (and thus the 
threshold value of $k_{\rm out}$ is also trivial), we found that otherwise the behavior 
is similar to the one shown in Fig.~\ref{n-t-sigma2}: When $k_{\rm out}$ converges from 
above towards the critical value $k_c = 2$, $\eta$ converges to a value that is larger 
than the value for short range critical SIR epidemics. Indeed this value seems to be 
compatible with the limit found for model (A), i.e.
\be
   \eta_c \equiv \lim_{k_{\rm out} \searrow k_c} \eta(k_{\rm out}) = 1.108(2)
\ee
for both model (A) and model (B) with $q=1/2$.

\begin{figure}
\includegraphics[width=0.53\textwidth]{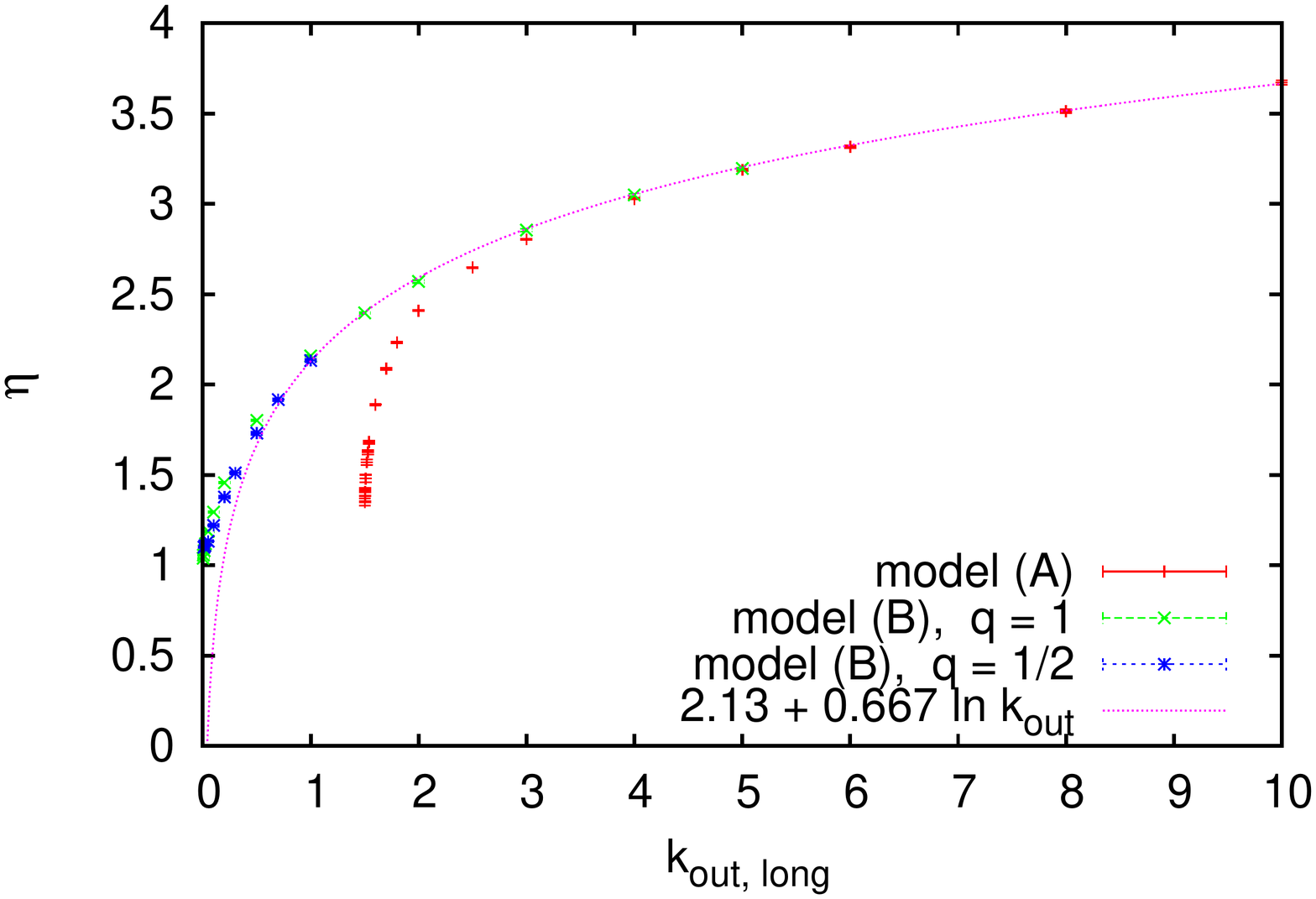}
\caption{(Color online) Plot of $\eta$ versus $k_{\rm out, long}$ for $\sigma=2$, where 
  $k_{\rm out, long}$ is equal to the number of long range contacts per node. For model (A) 
  it is equal to $k_{\rm out}$, while it is $k_{\rm out}-4q$ for model (B).  The 
  continuous curve indicates the leading behavior for large $k_{\rm out}$. In this limit,
  all three models have the same $\eta$.}
\label{sigma2-expon1}
\end{figure}

The exponents $\eta(k_{\rm out})$ are plotted against $k_{\rm out}$ in Fig.~\ref{sigma2-expon1},
together with analogous exponents for model (B).
 For large $k_{\rm out}$ one sees a logarithmic increase similar to the one 
found in one spatial dimension, 
\be
   \eta \approx 2.13 + 0.667 \ln k_{\rm out}.
\ee
Indeed, this behavior is common to both models, because short range bonds (which make the 
entire difference between models (A) and (B) are irrelevant when $k_{\rm out} \gg 1$.

\begin{figure}
\includegraphics[width=0.53\textwidth]{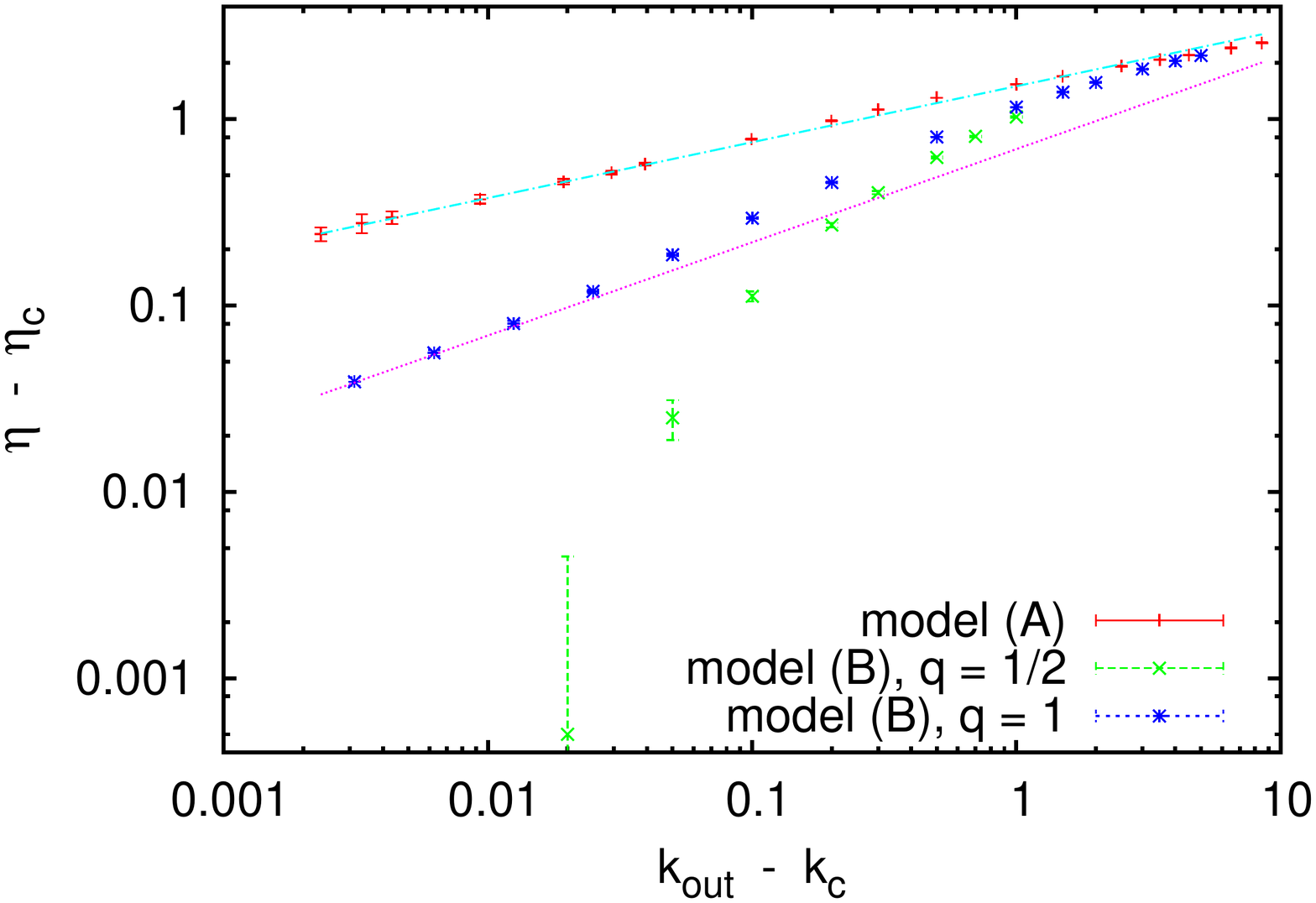}
\caption{(Color online) Log-log plot of the same data as in Fig.~\ref{sigma2-expon1}, with 
$k_{\rm out}-k_c$ on the horizontal axis and $\eta-\eta_c$ on the vertical axis. The straight 
lines correspond to power laws holding in the limit $k_{\rm out}-k_c \to 0$, with exponent 
0.3 for model (A) and exponent 1/2 for model (B) with $q=1$.}. 
\label{sigma2-expon2}
\end{figure}

The detailed threshold behavior of $\eta$ cannot be seen from Fig.~\ref{sigma2-expon1},
therefore we show in Fig.~\ref{sigma2-expon2} the same data plotted on a log-log plot, where 
we also changed the abscissa from $k_{\rm out}$ to $k_{\rm out}-k_c$ and the vertical axis from 
$\eta$ to $\eta=\eta_c$. We see clear indications
for power laws in model (A) and in model (B) with $q=1$, while the data suggest a different
behavior for model (B) with $q=1/2$. More precisely, the data for model (B) with $q=1$ 
suggest 
\be
   \eta -1 \sim (k_{\rm out}-4)^{1/2},
\ee
while for model (A) we find $\eta - \eta_c \sim (k_{\rm out}-k_c)^{0.3}$.

\begin{figure}
\includegraphics[width=0.53\textwidth]{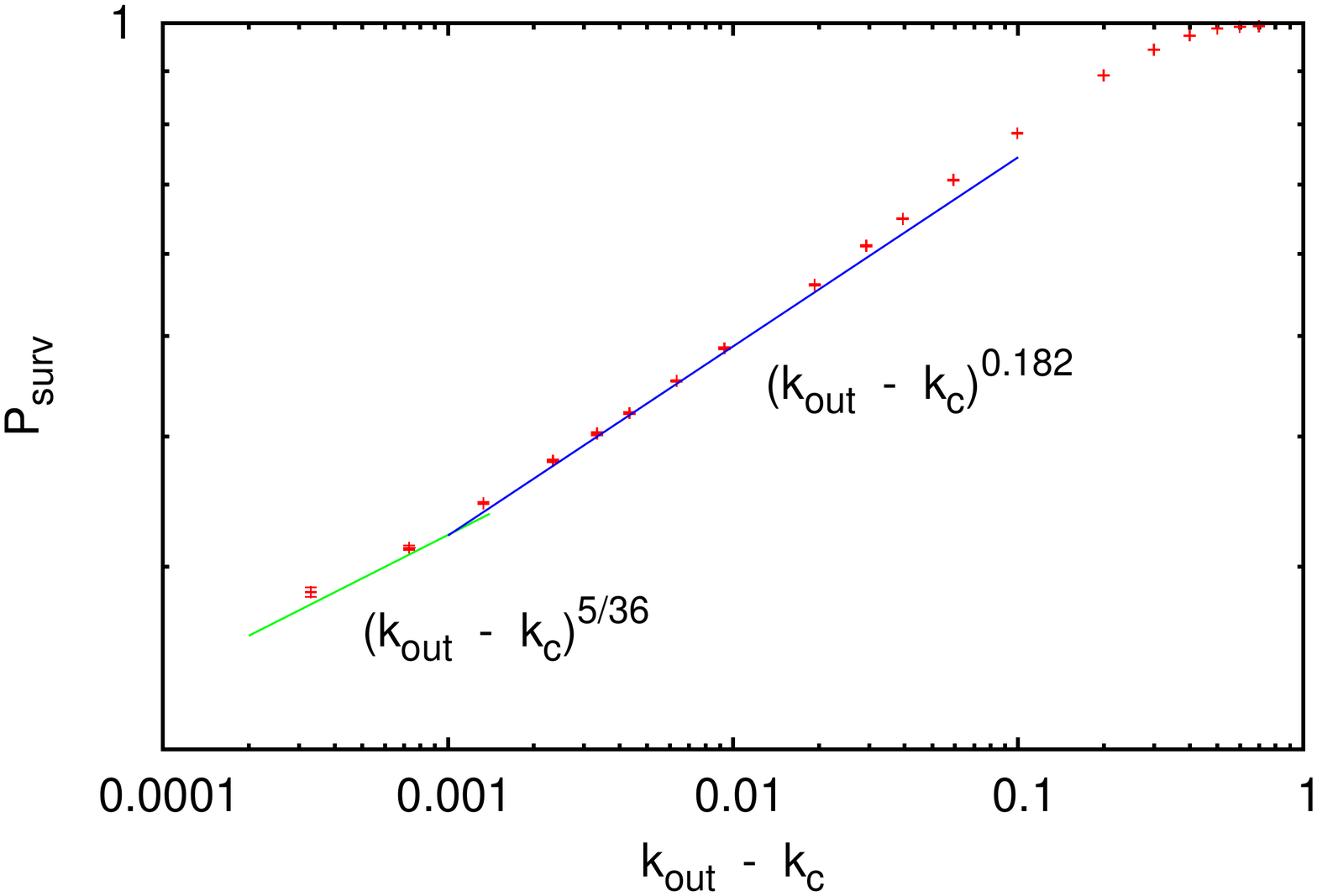}
\caption{(Color online) Log-log plot of the survival probability $P_{\rm surv}$ for model (A) 
   with $\sigma=2$.  The long straight line indicates that $P_{\rm survive}$ follows roughly 
   a power law with exponent $\approx 0.182$, but there are strong deviations. The short straight 
   line indicates the power law $P_{\rm surv} \sim (k_{\rm out}-k_c)^{0.14}$ that seems to hold very
   close to the critical point. The value of $k_c$ used in this plot is 0.50067, but the 
   plot is not very sensitive to the precise value.}
\label{Psurv}
\end{figure}

For $d=1$, the percolation transition at $\sigma = d$ is known to be 
discontinuous \cite{Aizenman,Aizenman88}, as found also numerically in I. This is not the 
case for $d=2$, as seen from Fig.~\ref{Psurv}. There we see that the survival probability 
$P_{\rm survive}$ vanishes at the critical point, as expected for a second order transition. 
Indeed, for values of $k_{\rm out}$ very close to $k_c$ we see that $P_{\rm survive} \sim (k_{\rm out}-k_c)^\beta$, 
where $\beta \approx 0.14$ is compatible with the order parameter exponent $\beta=5/36$ for ordinary 
(short range) critical percolation. This is not surprising, as we expect this scaling for 
all $\sigma >2$, but the huge corrections to scaling visible in Fig.~\ref{Psurv} 
indicate that we should be careful with any quick conclusion. The detailed behavior 
at the critical point will be discussed in the next section.

\section{Critical Behavior}

\subsection{``Easy" regions and general overview}

First we shall discuss the ``easy" regions, i.e. values of $\sigma$ far from the transition points
B, C and D in Fig.~1, while the vicinities of these points will be discussed in later 
subsections.

For $\sigma \ll 2/3$ one is far in the mean field regime. We do not show any data, but 
it suffices to say that all critical exponents agreed with with their predictions to 
very high accuracy. There was also no problem near point A in Fig.~1, where the 
{\it supercritical} model changes from mean field to the intermediate phase. We expect 
of course highly non-trivial behavior as one goes from the critical line into the 
supercritical phase, when $0 < \sigma < 2/3$ (see I for the analogous situation in $d=1$),
but this seems to leave no traces on the critical line.
Similarly, there is no problem for $\sigma \gg 2$, where we recover short range behavior. 

\begin{figure}
\includegraphics[width=0.53\textwidth]{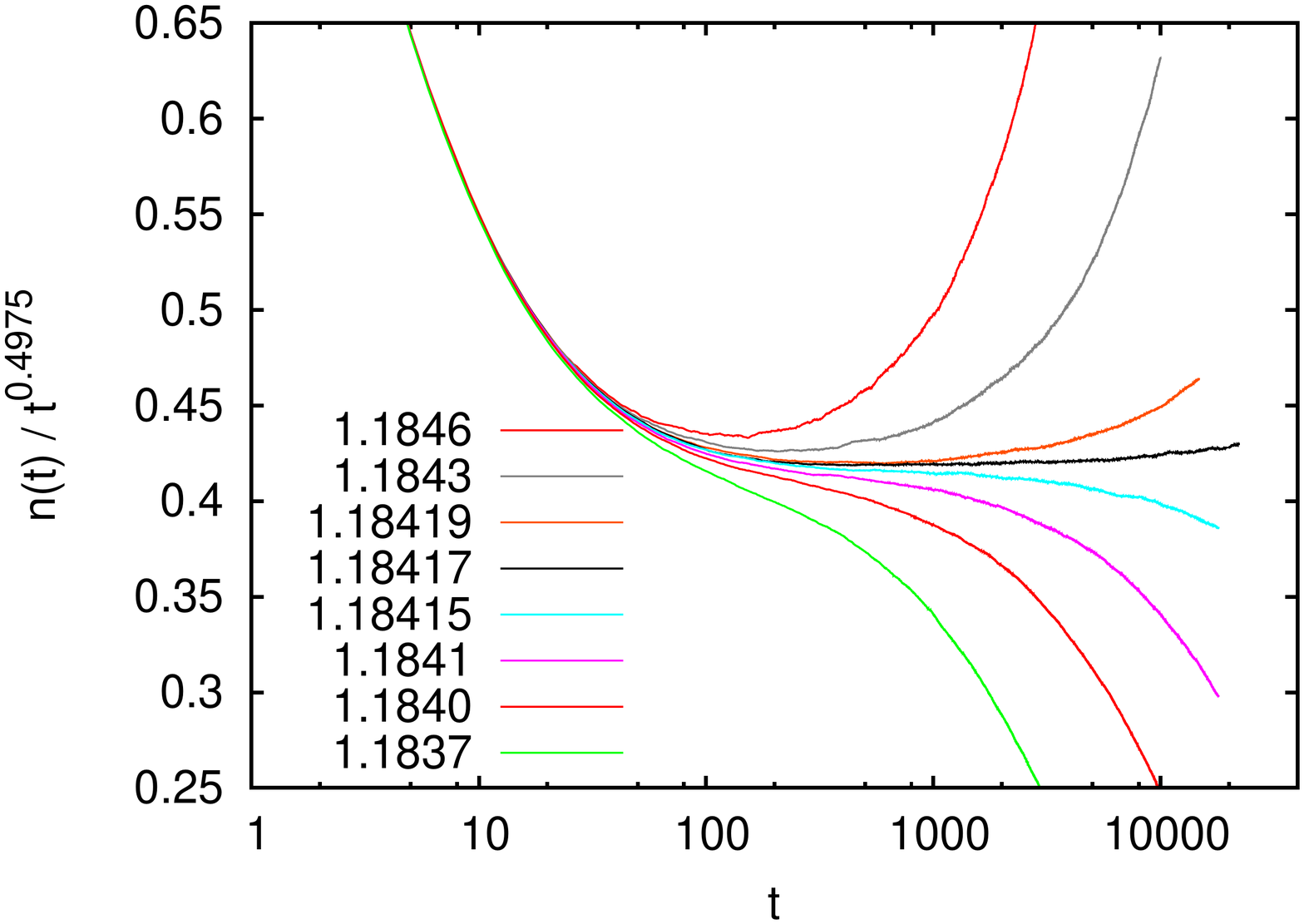}
\caption{(Color online) Log-linear plot of $t^{-\eta} n(t)$ for $\sigma = 1.25$, with $\eta = 0.4975$. 
   This value of $\eta$ was chosen as it seems to give the best horizontal extrapolation for $t\to
   \infty$.}
\label{N_crit}
\end{figure}

\begin{figure}
\includegraphics[width=0.53\textwidth]{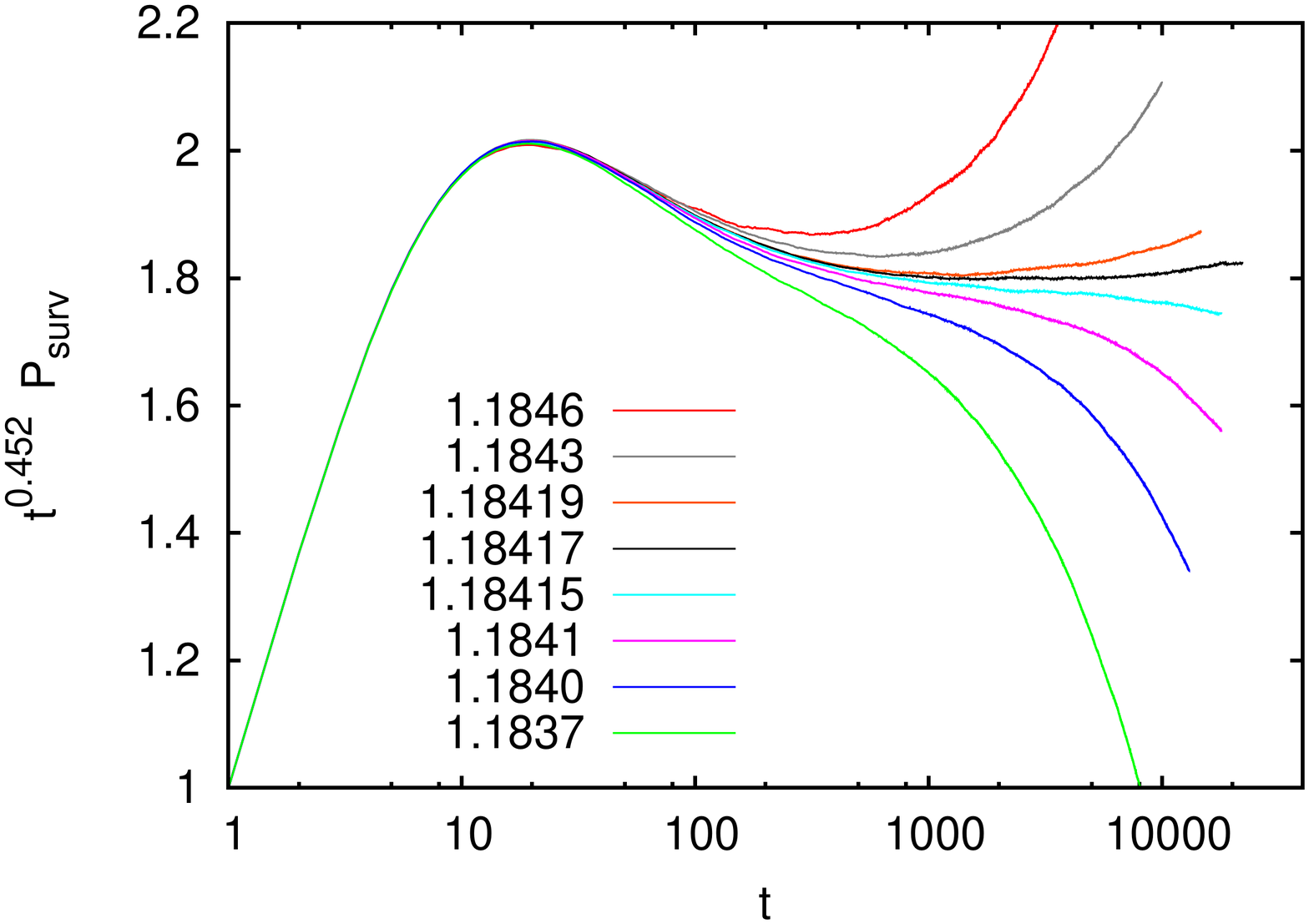}
\caption{(Color online) Log-linear plot of $t^\delta P_{\rm surv}(t)$ for $\sigma = 1.25$, with $\delta = 0.452$. 
   Again, the value of the exponent was chosen as it seems to give the best horizontal extrapolation for $t\to
   \infty$.}
\label{P_crit}
\end{figure}

\begin{figure}
\includegraphics[width=0.53\textwidth]{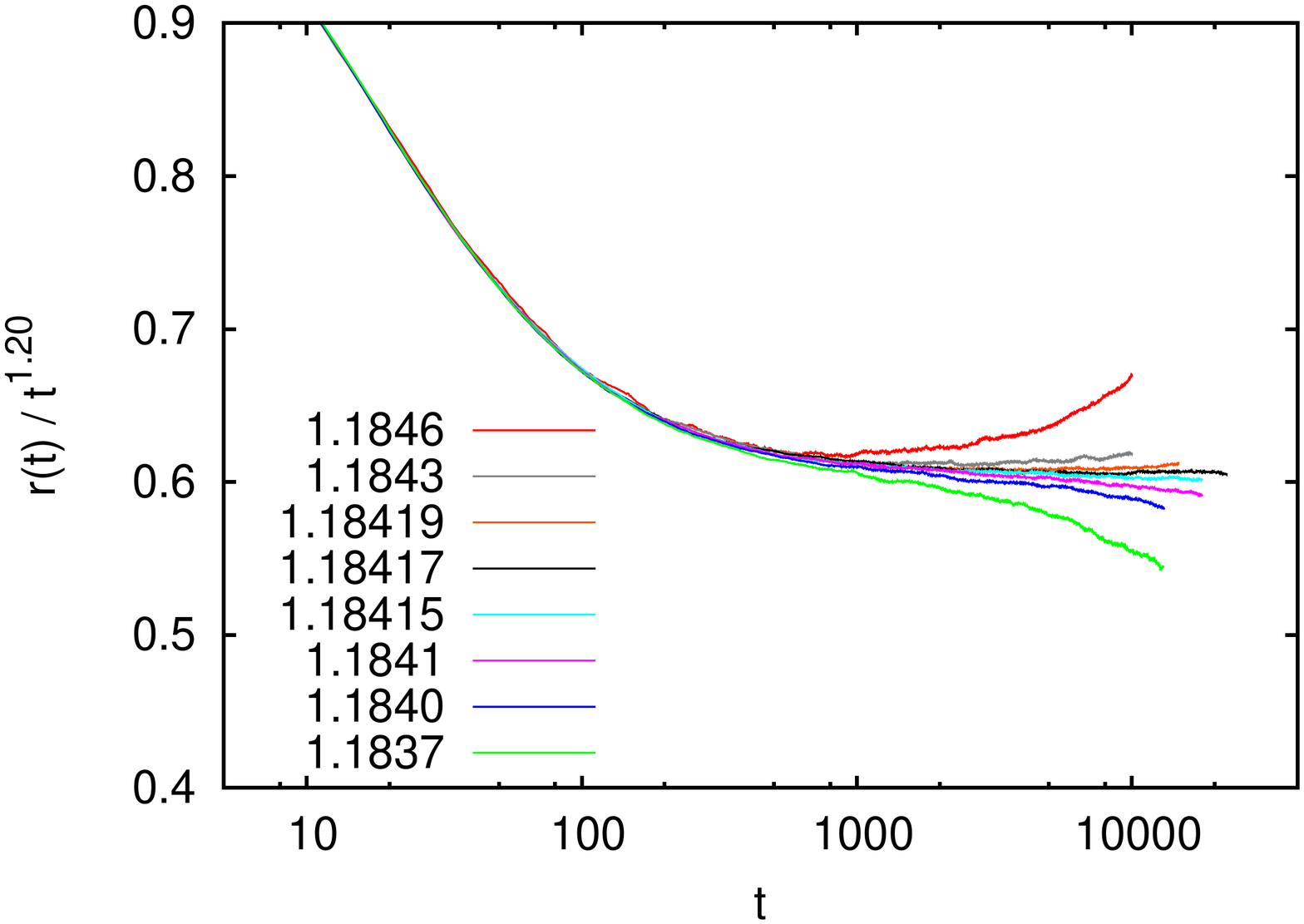}
\caption{(Color online) Log-linear plot of $t^{-1/z} r(t)$ for $\sigma = 1.25$, with $z = 0.833$. Here, $r(t)$
is the geometric average of the radius of the infected sites.}
\label{R_crit}
\end{figure}

\begin{figure}
\includegraphics[width=0.53\textwidth]{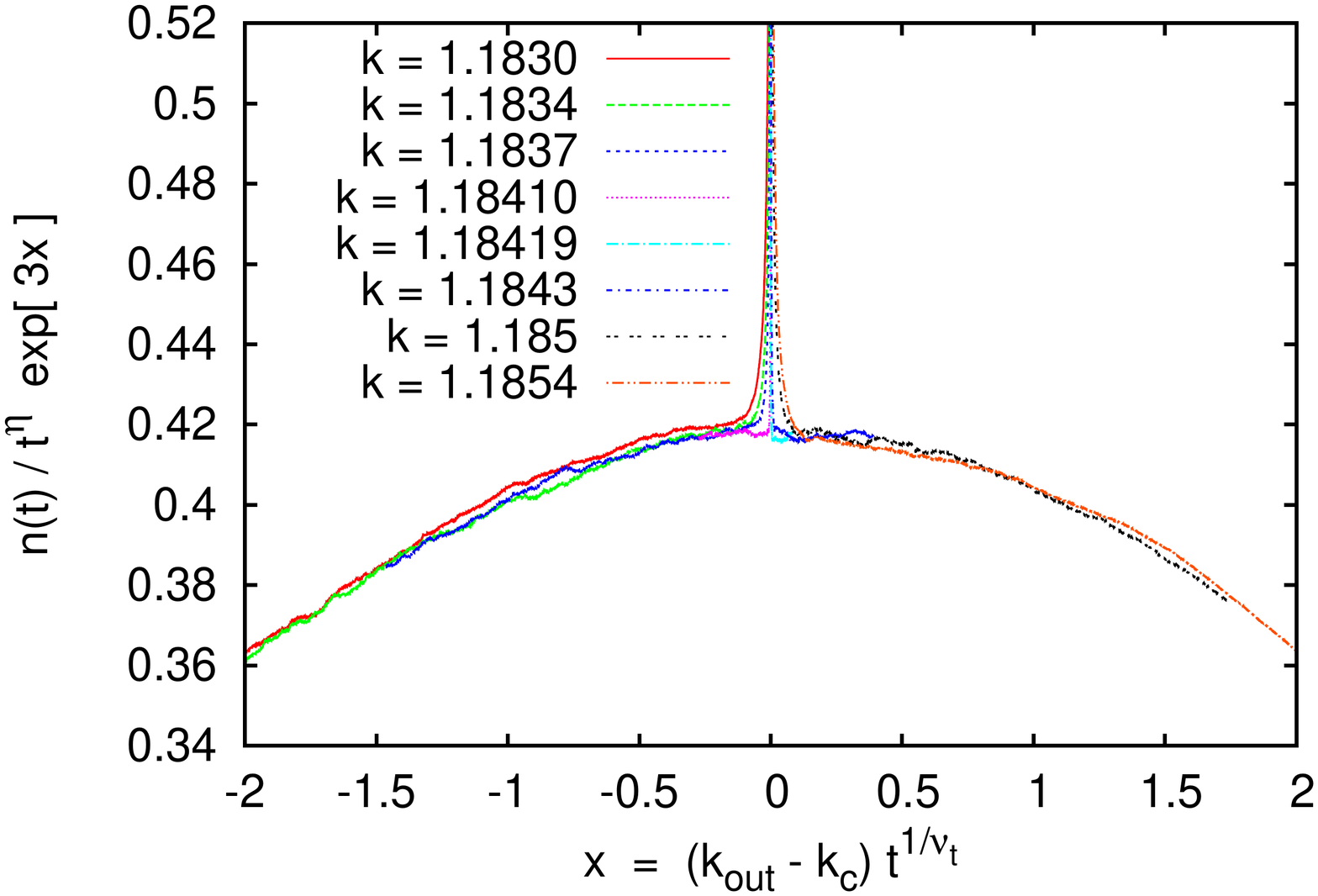}
\caption{(Color online) The quantity $t^{-\eta} n(t,\epsilon) e^{3x}$ plotted against $x \equiv \epsilon t^{1/\nu_t}$,
   for $\sigma = 1.25$. Here, $\epsilon = k_{\rm out}-k_c$, and $k_c$ and $\eta$ were as determined from
   Fig.~\ref{N_crit}. The factor $e^{3x}$ was added in order to make the scaling function horizontal 
   at the origin and increasing thereby the visible resolution.}
\label{collaps-13}
\end{figure}

\begin{figure}
\includegraphics[width=0.53\textwidth]{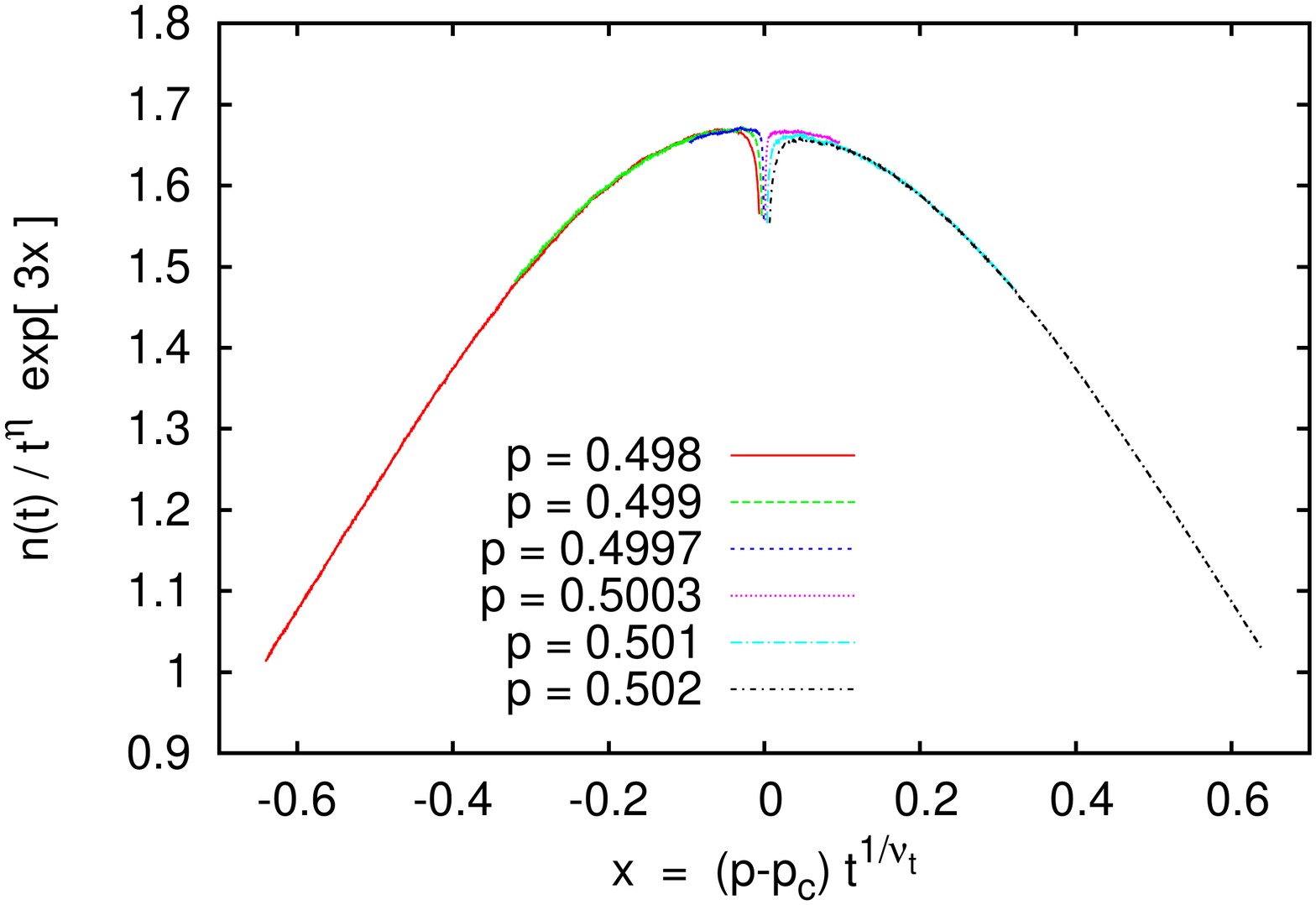}
\caption{(Color online) Plot analogous to Fig.~\ref{collaps-13} but for ordinary bond percolation.
   Here we used the exact value $p_c=1/2$ and the critical exponent estimates from \cite{Grass-1992}.}
\label{collaps-bp}
\end{figure}

\begin{figure}
\includegraphics[width=0.53\textwidth]{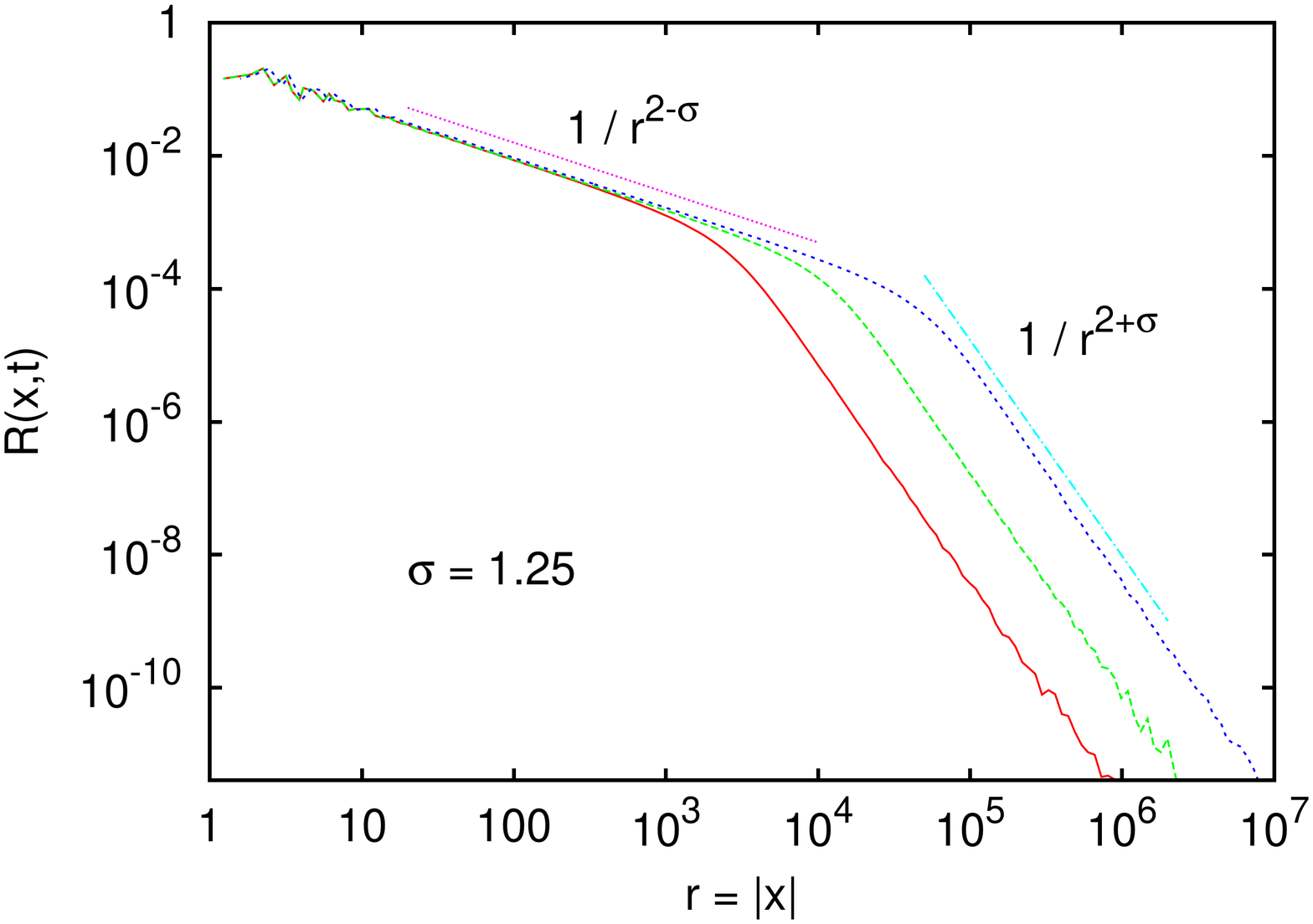}
\caption{(Color online) Log-log plots of densities of removed sites for $\sigma=1.25$ at 
   $t=2000, 7000,$ and 22000. As in Figs.~\ref{front-1.60} and \ref{front-0.91}, the 
   fluctuations at small $r$ are artifacts. The straight lines indicate the power laws for 
   small and large $r$.} 
\label{front-1.3}
\end{figure}

For intermediate values $2/3 < \sigma < 2$ we do not have exact predictions for the 
critical exponents, but in the central part of this region, say $0.9 < \sigma < 1.4$, 
we find rather clean scaling laws with only moderate corrections to scaling. Typical 
results obtained for $\sigma = 1.25$, e.g., are shown in Figs.~\ref{N_crit} to \ref{front-1.3}.

In Fig.~\ref{N_crit} we show the average number of infected sites as a function of $t$
for various values of $k_{\rm out}$ near the critical point. More precisely, in order 
to do justice to the very small error bars, we plotted $n(t) / t^\eta$, where $\eta$
is an estimate for the critical exponent. If it is chosen correctly, the data for 
$k_{\rm out}=k_c$ should be horizontal for large $t$. We see of course considerable 
corrections for $t < 200$, but they do not prevent precise estimates of $\eta$ and $k_c$.

Similar plots for the survival probability and for the (geometric) average radius of 
the cluster of infected sites are shown in Figs.~\ref{P_crit} and \ref{R_crit}. They 
show that power laws 
\be
   P_{\rm surv}(t) \sim t^{-\delta}\;, \qquad r(t) \sim t^{1/z}
\ee
are indeed observed for the same value of $k_{\rm out}$. Also they 
satisfy the hyperscaling relation $\eta = d/z-2\delta-1$.

While $\eta, \delta,$ and $z$ are sufficient to describe scaling exactly {\it at} the 
critical point, one more exponent is needed to describe scaling {\it near} the critical point. 
For this we can use e.g. the order parameter exponent $\beta$ or the correlation length exponent $\nu$.
For our growth simulations, the most convenient exponent is, however, $\nu_t$ which describes 
how the correlation {\it time} diverges as $k_{\rm out} \to k_c$. It also describes how fast the 
different curves in Figs.~\ref{N_crit} to \ref{R_crit} diverge at $t\to\infty$.
Technically it is defined by defining first $\epsilon = k_{\rm out}-k_c$, and using then the 
finite-time scaling {\it ansatz} 
\be
   n(t,\epsilon) \approx t^{\eta} F(\epsilon t^{1/\nu_t}),
\ee
where $F(z)$ is an everywhere analytic universal scaling function. 

The most precise way to measure it in ordinary percolation is to relate $\partial 
n(t,\epsilon)/\partial \epsilon$ to correlations between cluster sizes and cluster perimeter 
lengths \cite{Lorenz-Ziff,grass-highdim,grass-tricrit}. In the present case this cannot be used, 
so we had to estimate it from data collapse plots, although these are notoriously unreliable 
in the presence of strong corrections to scaling. Results for $\sigma = 1.25$ are shown in 
Fig.~\ref{collaps-13}, and analogous results for ordinary bond percolation are shown for comparison 
in Fig.~\ref{collaps-bp}. In order to improve the accuracy, we plotted in both cases not 
$t^{-\eta} n(t,\epsilon)$ against $x \equiv \epsilon t^{1/\nu_t}$, but we multiplied the former 
by a suitable exponential $e^{ax}$, where the constant $a$ is chosen such as to make the 
scaling function $F(z)$ horizontal at $z=0$.

We see a good data collapse both in Fig.~\ref{collaps-13} and in Fig.~\ref{collaps-bp}, with
obvious deviations near $z=0$ due to scaling violations at small $t$ (in Fig.~\ref{collaps-bp}
we only plotted data for $t>5$). For bond percolation we used of course the exactly known 
values for $p_c$ and the estimates of the critical exponents from \cite{Grass-1992}, while 
we used in Fig.~\ref{collaps-13} the 
parameters determined from Fig.~\ref{N_crit}. We obtained $\nu_t = 1.176(14)$.  

Finally, let us discuss the densities of removed sites, i.e. the densities of the percolation
cluster at different times (the densities of active sites are trivially related). 
In Fig.~\ref{front-1.3} we see two different power laws (as also found in I for the 1-d case). 
More precisely, we find the same scaling law Eq.~(\ref{super-dens}), but with 
\be
   \phi(z) = \left\{
            \begin{array}{rl}
            z^{-2+\sigma}  & \text{ for } z \ll 1,\\
            z^{-2-\sigma}  & \text{ for } z \gg 1 ,
            \end{array} \right.
     \label{crit-dens1}
\ee
Again $\xi(t)$ can be taken equal to $r(t)$, the geometric average cluster radius. As in 
the 1-d case this gives immediately a relation between the critical exponents \cite{Janssen99},
\be
    (1+\eta)z = \sigma
\ee
(notice that $z$ was defined differently in I, as $r\sim t^z$, while we now use the more conventional
definition $t\sim r^z$). This is satisfied within statistical errors.

\begin{figure}
\includegraphics[width=0.53\textwidth]{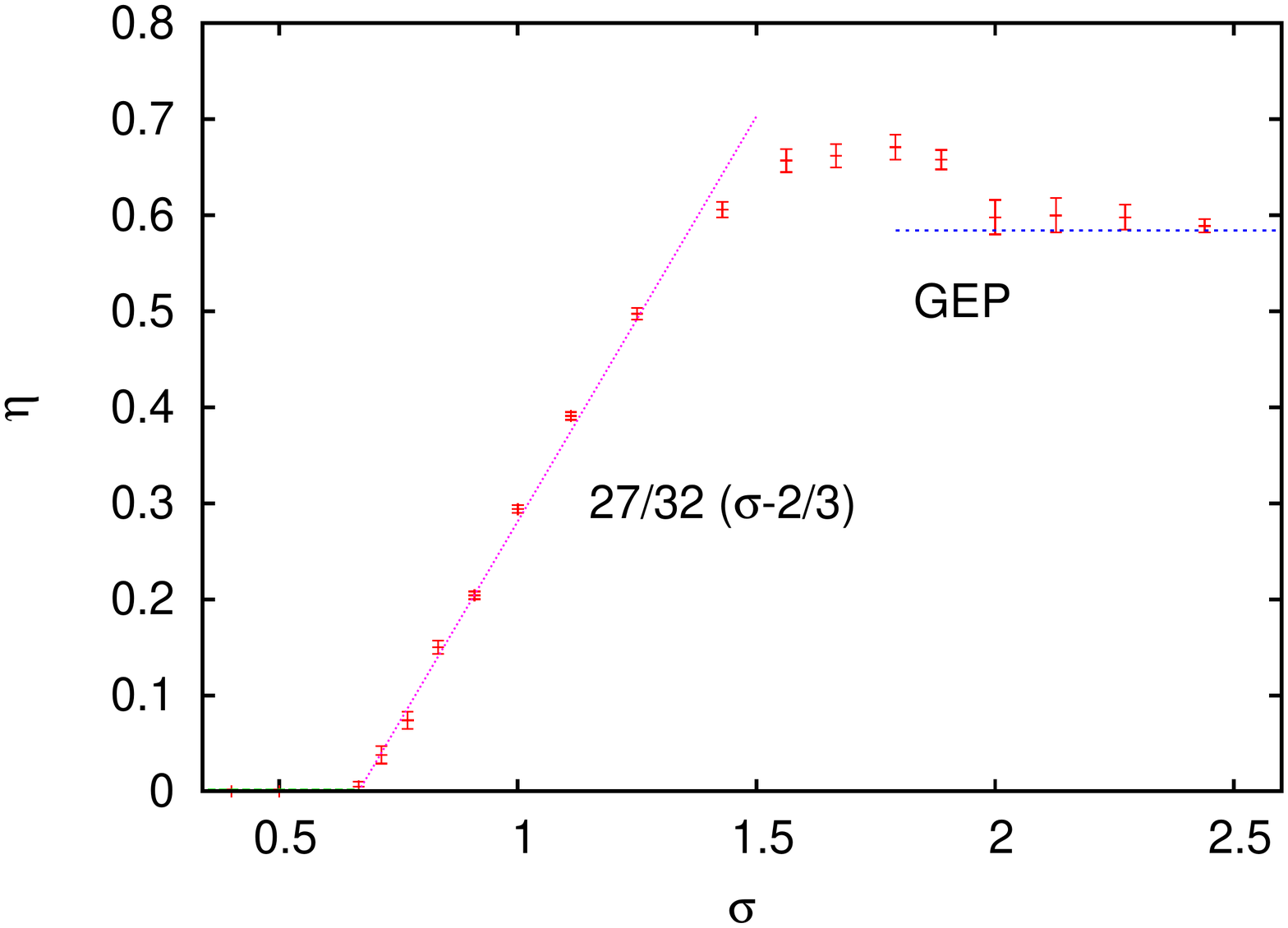}
\caption{(Color online) The growth exponent $\eta$ plotted against $\sigma$. For $\sigma \leq 2/3$,
   the mean field prediction is $\eta=0$. For ordinary percolation (``GEP") the result of \cite{Grass-1992}
   is $\eta =0.58435(50)$. The leading order $\epsilon$-expansion result is as indicated by the 
   tilted straight line.}
\label{eta.fig}
\end{figure}

\begin{figure}
\includegraphics[width=0.53\textwidth]{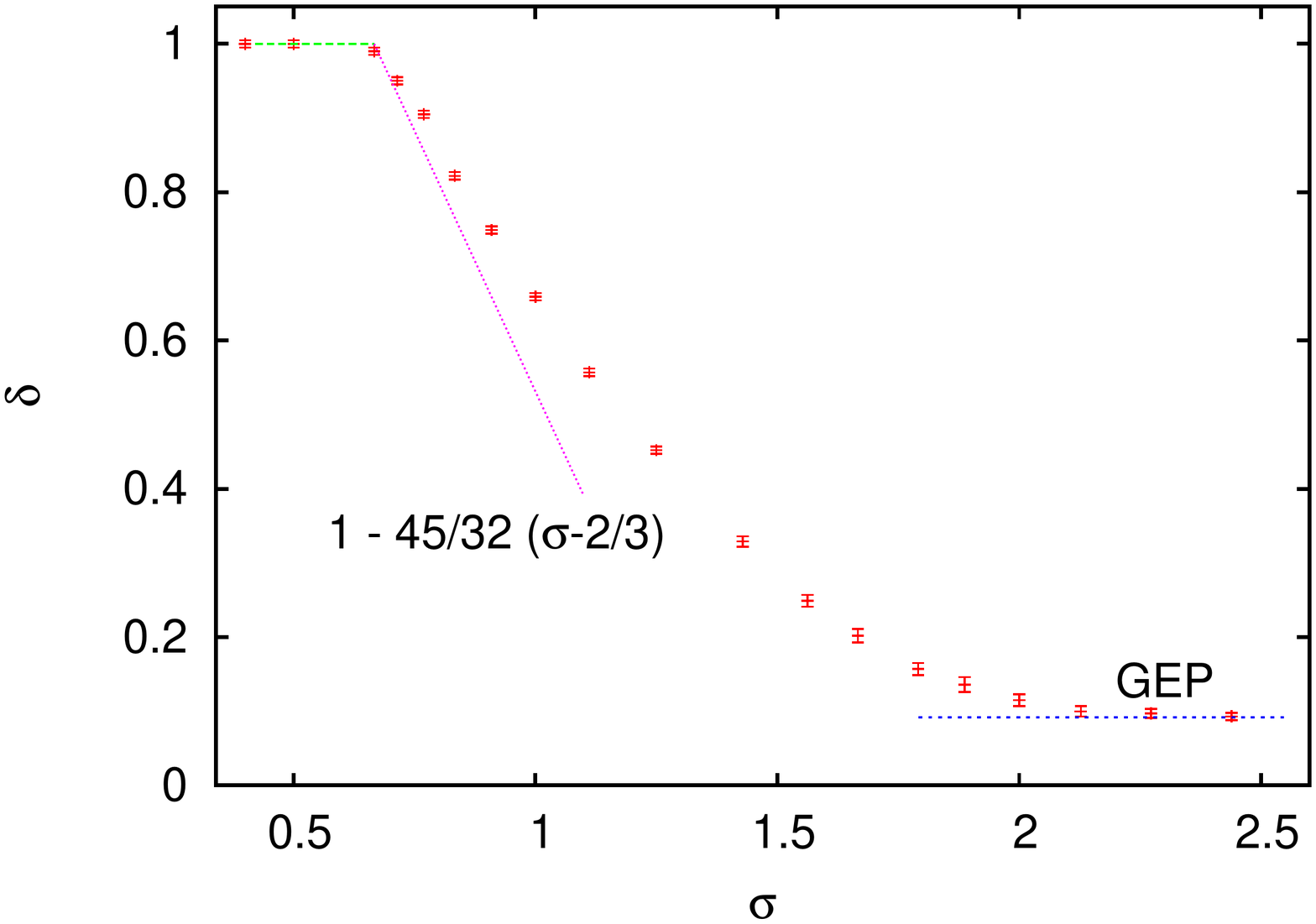}
\caption{(Color online) The survival exponent $\delta$ plotted against $\sigma$. For $\sigma \leq 2/3$,
   the mean field prediction is $\delta=1$. For ordinary percolation (``GEP") the result of \cite{Grass-1992}
   is $\delta = 0.09211(3)$.}
\label{delta.fig}
\end{figure}

\begin{figure}
\includegraphics[width=0.53\textwidth]{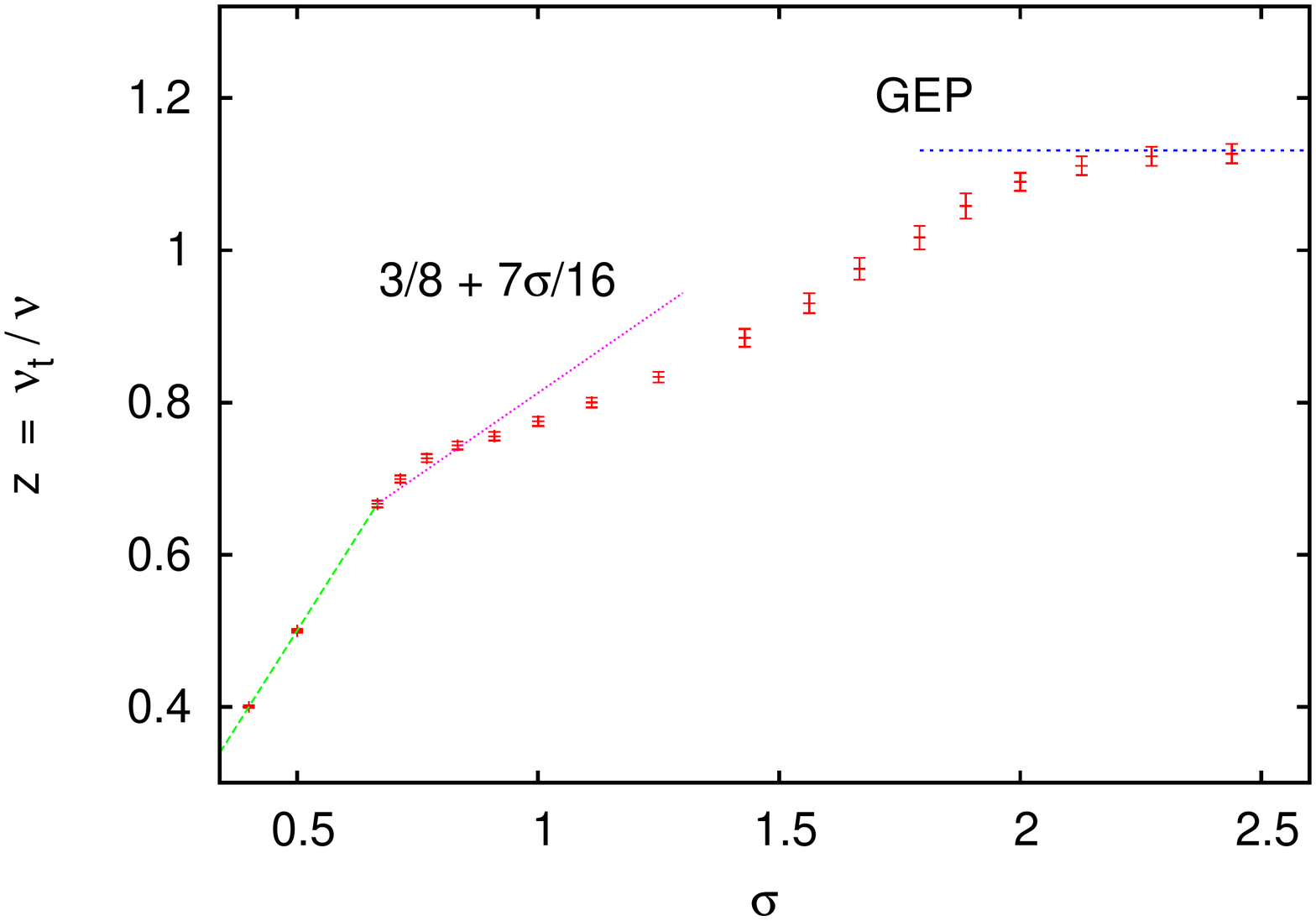}
\caption{(Color online) The dynamical exponent $z$. For $\sigma \leq 2/3$,
   the mean field prediction is $z=\sigma$, while the GEP result is 1.1309(4).}
\label{z.fig}
\end{figure} 

\begin{figure}
\includegraphics[width=0.53\textwidth]{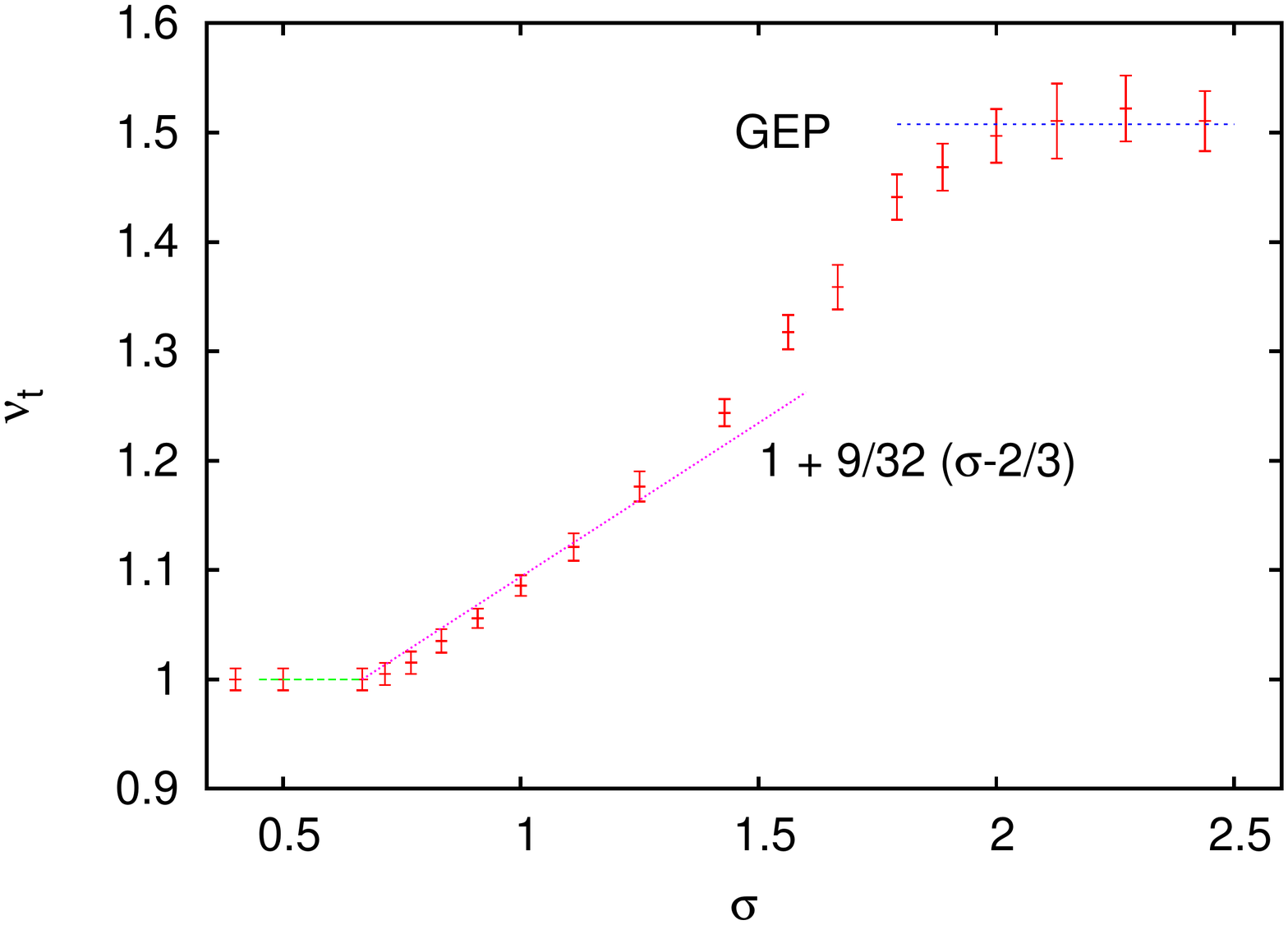}
\caption{(Color online) The correlation time exponent $\nu_t$ obtained by collapse plots like 
Fig.~\ref{collaps-13}. The mean field result is $\nu_t=1$, and the GEP result is 1.5078(5).}
\label{nu_t.fig}
\end{figure} 

\begin{figure}
\includegraphics[width=0.53\textwidth]{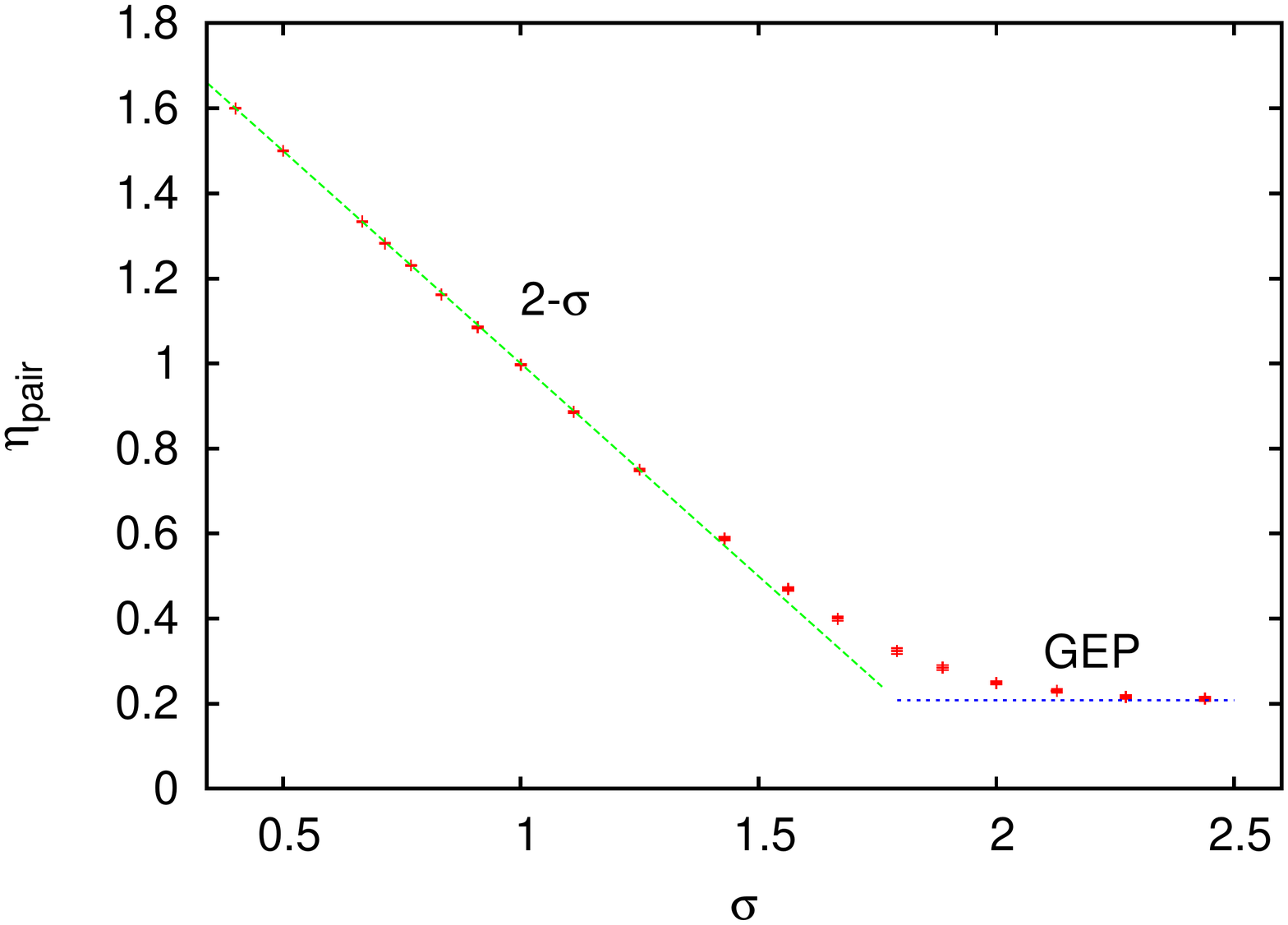}
\caption{(Color online) The pair connectedness exponent $\eta_{\rm pair}$. According to 
\cite{Janssen99,Linder} it is exactly equal to $2-\sigma$ for all values of $\sigma$ up to 
the point where it assumes the GEP value $\eta_{\rm pair}=2\beta/\nu = 0.2083$.} 
\label{pair.fig}
\end{figure}

\begin{figure}
\includegraphics[width=0.53\textwidth]{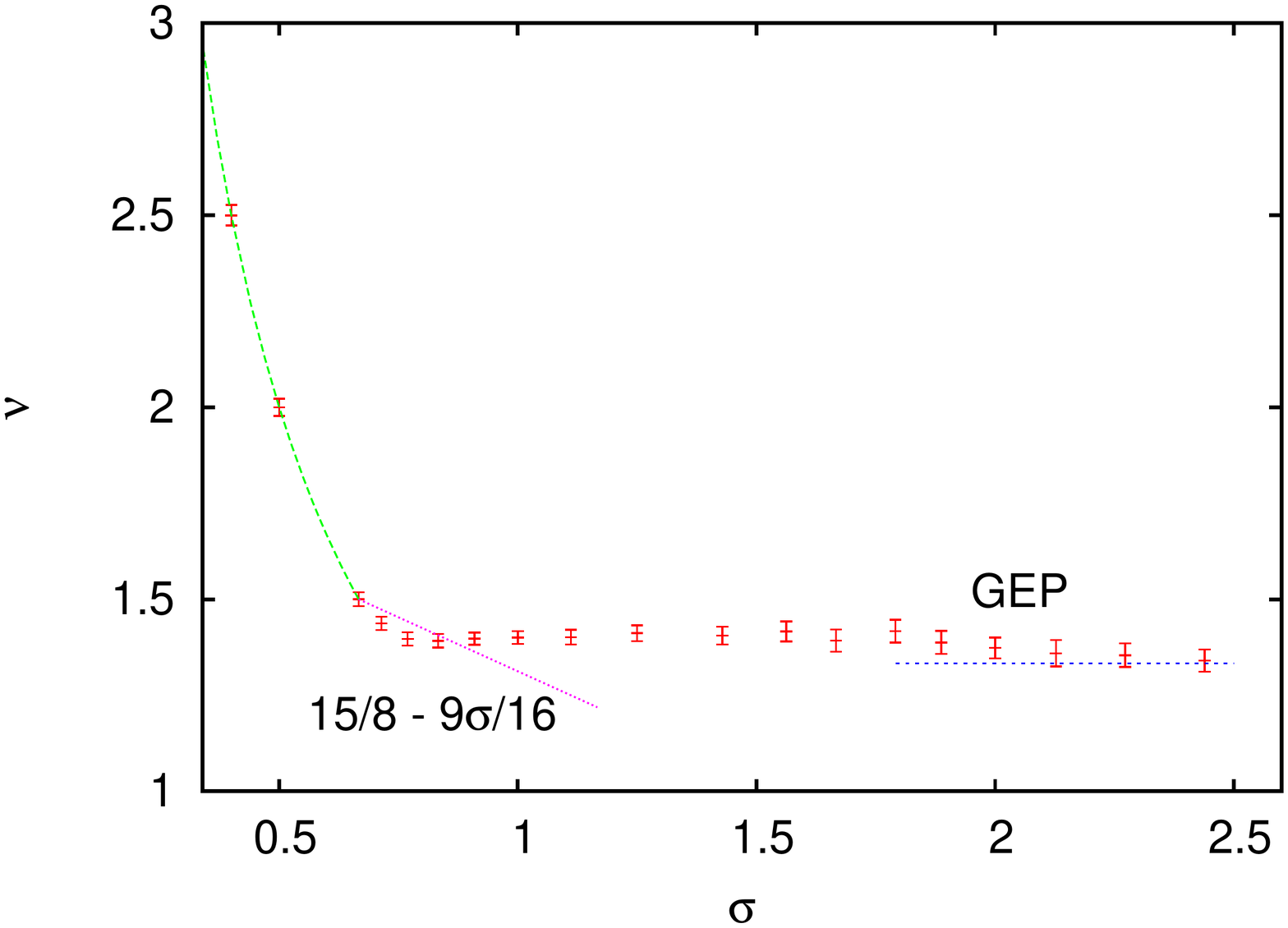}
\caption{(Color online) The correlation length exponent $\nu$ obtained by $\nu = \nu_t/z$. 
We include it, and the exponent $\beta$ in the next figure, as the only exponents that are not 
measured directly. The GEP value is $\nu=4/3$.}
\label{nu.fig}
\end{figure}

\begin{figure}
\includegraphics[width=0.53\textwidth]{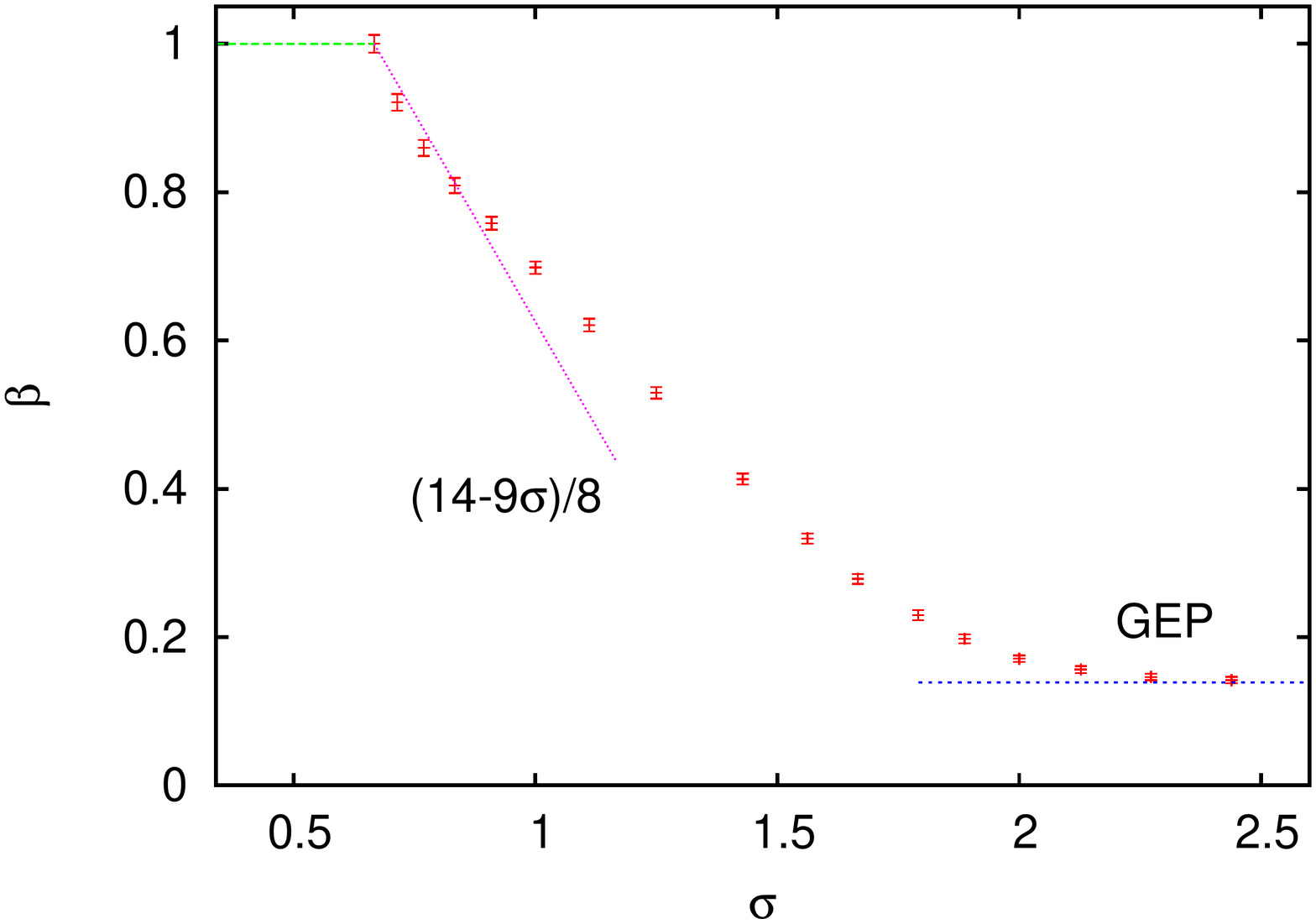}
\caption{(Color online) The order parameter exponent $\beta$ obtained by the hyperscaling relation
   $\beta = \nu\eta_{\rm pair}/2$. The GEP value is $\nu=5/36$. In the mean field regime $\beta=1$, 
   violating there the hyperscaling relation.}
\label{beta.fig}
\end{figure}

In general, we define the pair connectedness exponent $\eta_{\rm pair}$ by 
\be
   R({\bf x},t) \sim r^{-\eta_{\rm pair}}
\ee
for $r < r(t)$. We also obtain the correlation length exponent $\nu$ by $\nu = \nu_t/z$. Using 
these, we summarize our numerical results for the critical exponents in Figs.~\ref{eta.fig} to
\ref{beta.fig}. In these plots we also indicate the results for ordinary percolation (the 
``General Epidemic Process, GEP) and for mean percolation, and the $\epsilon$-expansion results
of \cite{Janssen99}. While these results need no further comment in the ``easy" regions, the 
results in the ``hard" regions will be discussed in the following subsections. Here we just point 
out that all exponents are very precisely given by the mean field values for $\sigma < 3/2$.
For $\sigma >2$ there are much larger corrections to the (presumably exact) GEP values, 
hinting at considerable finite-time corrections. We should say that the error bars in the 
``hard" regions are dominated by uncertainties in the extrapolation $t\to\infty$. They are 
not straightforward statistical errors, and their estimation is highly subjective, as for 
all critical exponents. It is quite obvious that some error bars (e.g. those for $\sigma >2$)
are wrong, but we included them on purpose, stressing thereby that meaningful error bars
are virtually impossible. As we said, details are given in the following subsections.

In the intermediate region, different 
exponents follow the predictions of the $\epsilon$-expansion to varying degree. 
Overall, the agreement is best for $\eta_{\rm pair}$, while it is worst for $z$.

\subsection{The critical case for $\sigma \approx 2/3$}

\begin{figure}
\includegraphics[width=0.53\textwidth]{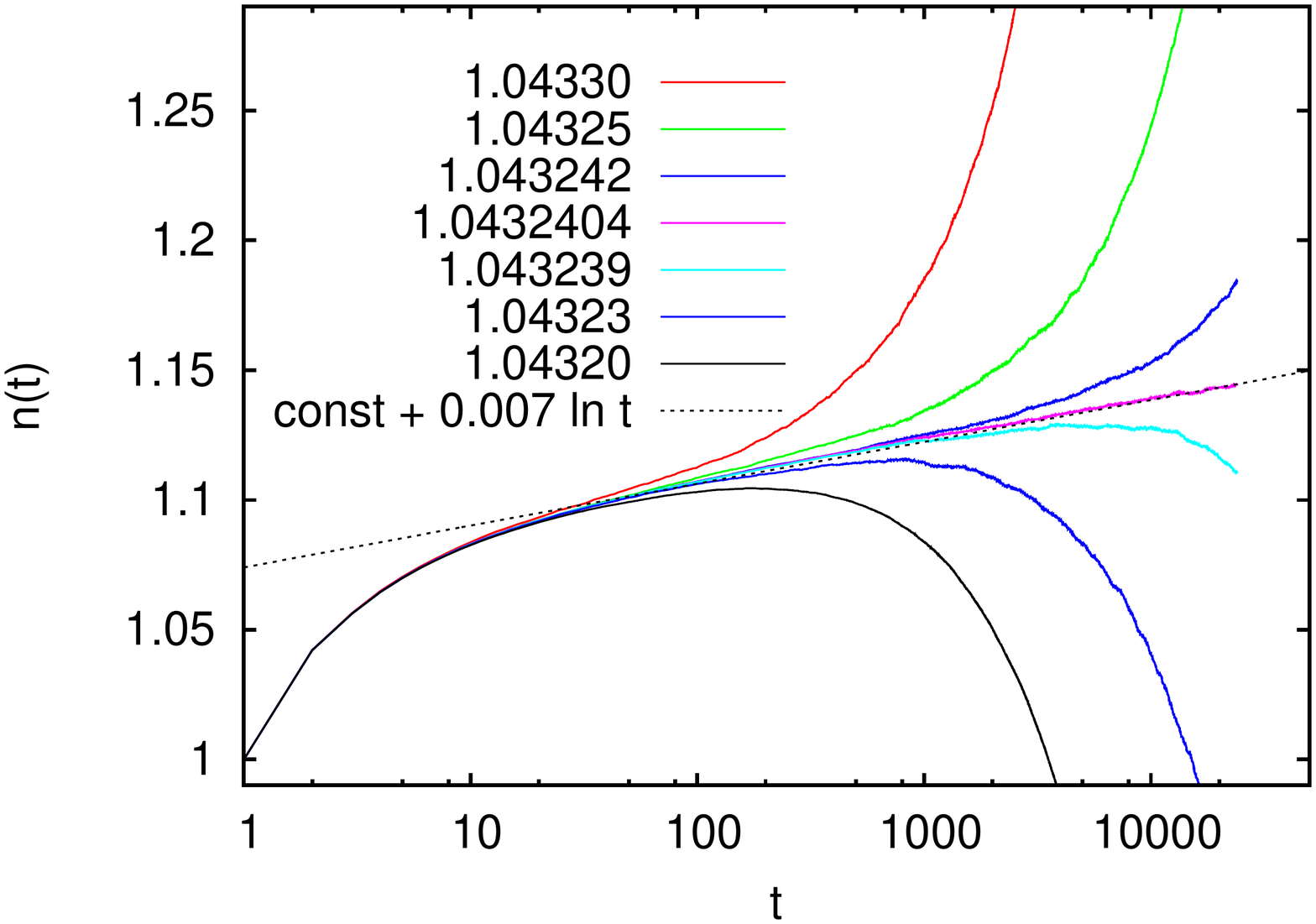}
\caption{(Color online) Log-linear plots of $ n(t)$ for $\sigma = 2/3$, for several values 
   of $k_{\rm out}$ near $k_c$. The central (near-horizontal) curve is obtained from $\approx 10^9$
   clusters, accordingly its statistical errors are smaller than the line thickness. The dotted 
   straight line is one (but not the preferred) possibility for logarithmic corrections. }
\label{N_crit-mf}
\end{figure}

As we said, we encountered no problems for $\sigma < 2/3$. Results exactly at $\sigma=2/3$
are shown in Fig.~\ref{N_crit-mf}. We see that indeed $\eta = 0$ at the critical point, with 
very small but clearly visible corrections. These corrections are compatible with being 
logarithmic,
\be
   n(t,k_c) \sim 1.074 + .007 \ln t,
\ee
but the smallness of the amplitude suggests that they are more likely to be log-log corrections,
\be
   n(t,k_c) \sim const+a \ln \ln t
\ee
with $a \approx 0.1$. Precise fits of this form are if course meaningless, unless they were 
guided by theory.

\begin{figure}
\includegraphics[width=0.53\textwidth]{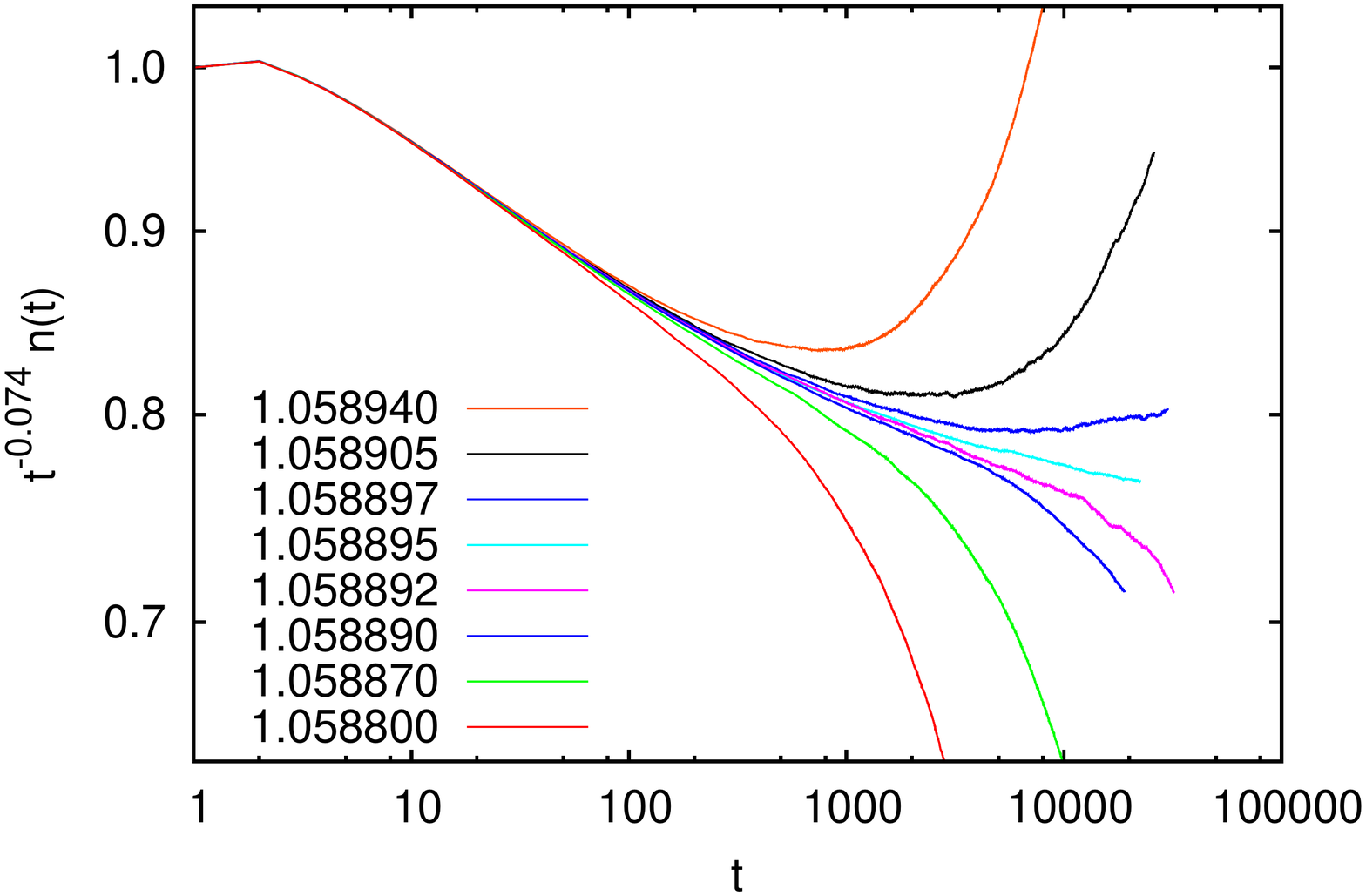}
\caption{(Color online) Log-log plots of $ t^{-\eta}n(t)$ for $\sigma = 0.7692$ and $\eta = 0.074$,
   for several values of $k_{\rm out}$ near $k_c$. None of the curves is linear, showing that the scaling 
   limit is not yet reached even for $t=10^4$. But curves for $k_{\rm out}\leq 0.058892$ 
   finally turn upward, showing that $k_c$ is smaller than this value.}
\label{N_crit-18}
\end{figure}

The situation is much worse for $\sigma$ slightly larger than 2/3, as also seen from 
Figs.~\ref{eta.fig} to \ref{beta.fig}. In Fig.~\ref{N_crit-18} we show $t^{-0.074}n(t)$ for 
$\sigma = 0.7692$. None of the curves in this plot is linear, showing that the scaling
limit is not yet reached even for $t=10^4$. Curves for $k_{\rm out}\leq 0.058892$ finally turn 
upward, showing that $k_c$ is larger than this value. If we assume conservatively that the 
curve for $k_{\rm out}=0.058895$ becomes asymptotically horizontal, then the value $\eta=0.074$
used in this plot is the correct exponent. But it is smaller than the value 0.087 predicted by
the $\epsilon$-expansion, indicating that corrections to scaling are even larger than suggested 
by Fig.~\ref{N_crit-18}. Similar problems were seen also for $\sigma = 0.714$ and 0.833, and had
been found also in I for $d=1$.

\subsection{The intermediate to short range cross-over}

Even worse corrections to scaling were found at the cross-over from intermediate to short 
range behavior, near points C and D in Fig.~1.

\begin{figure}
\includegraphics[width=0.53\textwidth]{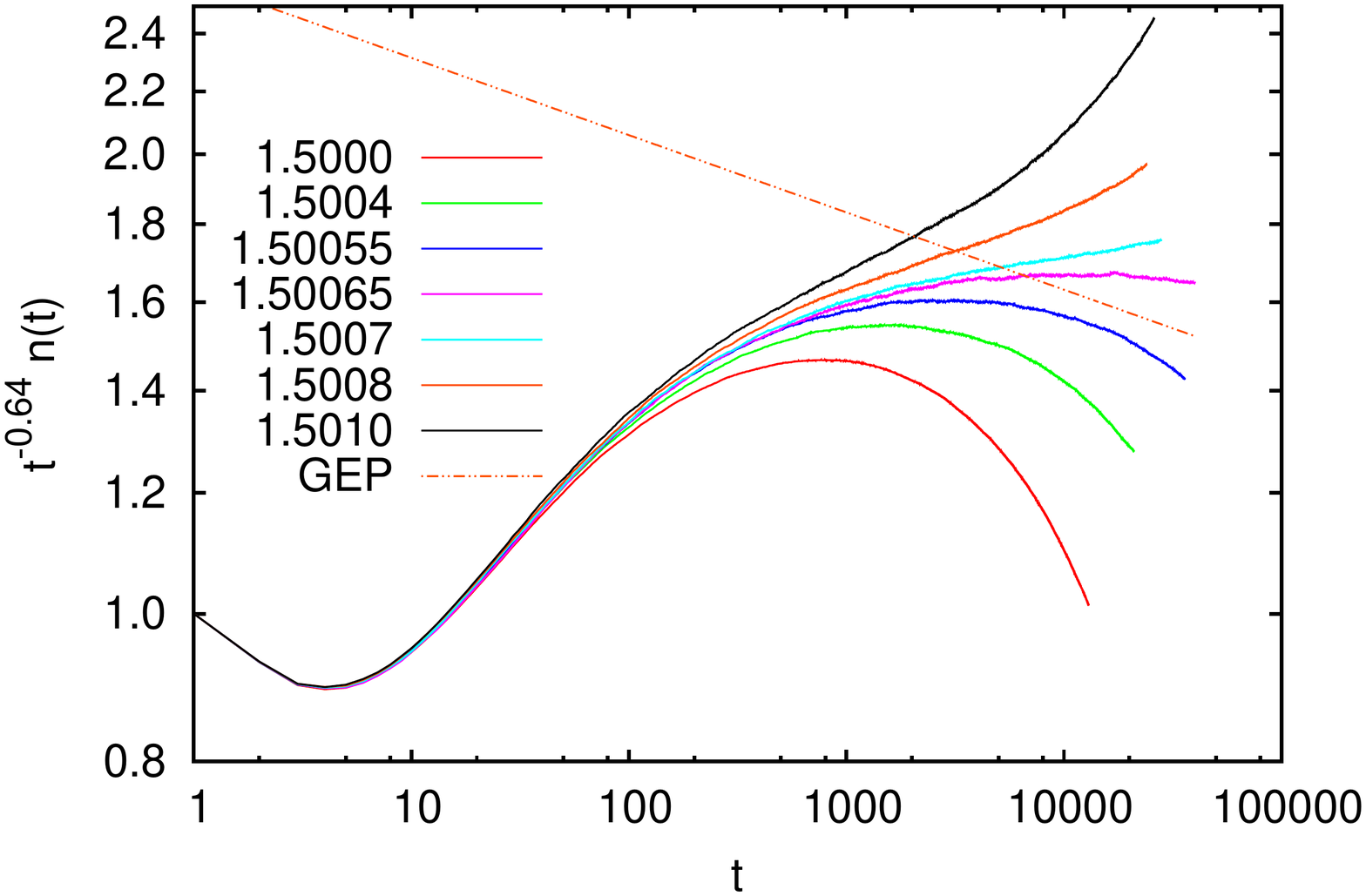}
\caption{(Color online) Log-log plots of $ t^{-0.64} n(t)$ for $\sigma = 2$, for several values
   of $k_{\rm out}$ near $k_c$. None of the curves is straight for any extended interval of $t$.
   The dashed-dotted straight line has the slope expected for the GEP, i.e. for short-range
   infection. The central (near-horizontal) curve is obtained from $\approx 10^9$
   clusters, accordingly its statistical errors are smaller than the line thickness. The dotted
   straight line is one (but not the preferred) possibility for logarithmic corrections. }
\label{N_crit-sigma2}
\end{figure}

\begin{figure}
\includegraphics[width=0.53\textwidth]{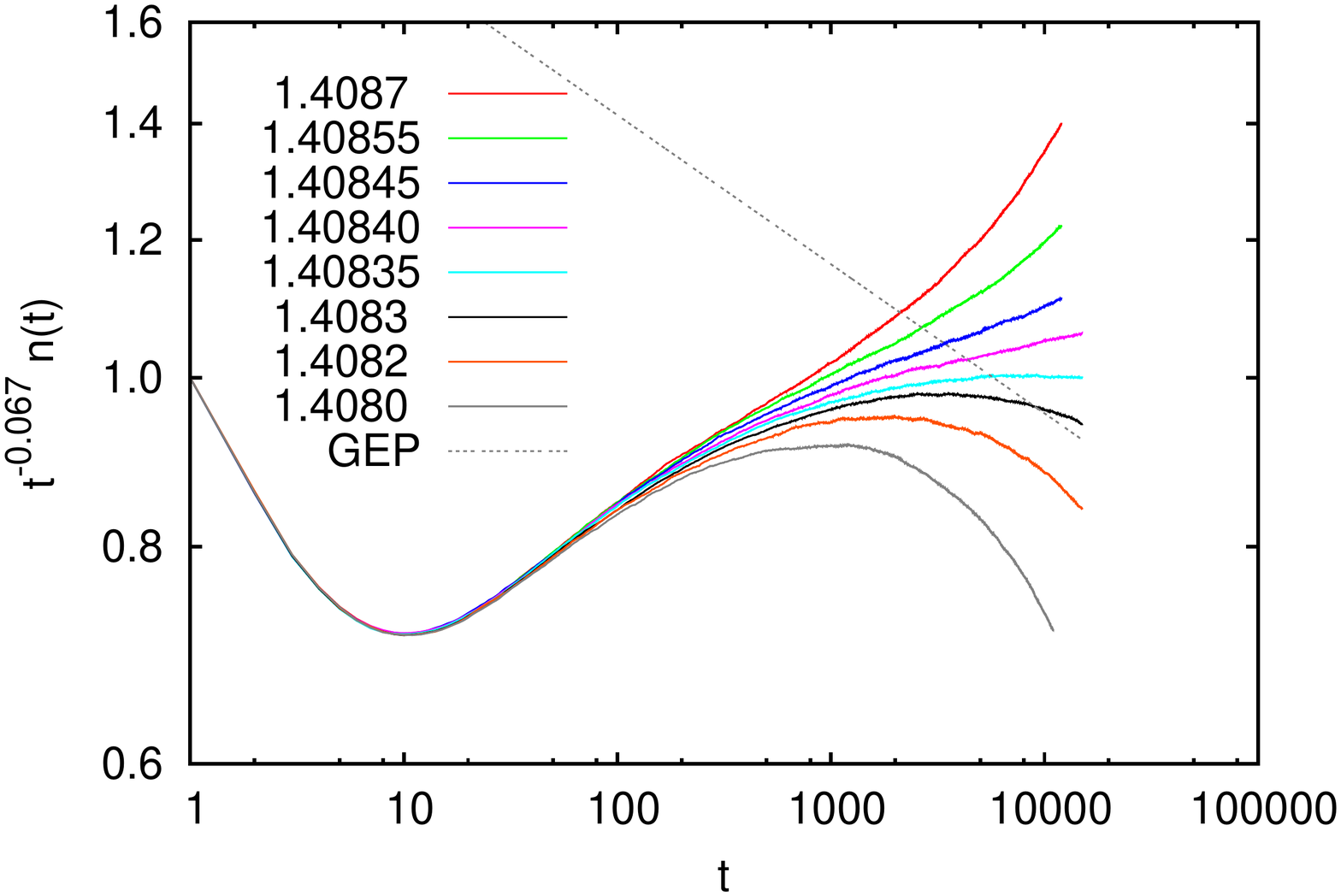}
\caption{(Color online) Plot analogous to Fig.~\ref{N_crit-sigma2}, but for 
   $\sigma=43/24 = 1.7916\ldots$ and with a different prefactor on the y-axis.}
\label{N_crit-sigmaC}
\end{figure}

The growth of $n(t)$ at $\sigma=2$ is shown in Fig.~\ref{N_crit-sigma2}. More precisely, we 
show there $t^{-0.64} n(t)$, where the power of the prefactor was chosen so that the most 
straight curves are roughly horizontal at large $t$. But none of the curves is really straight. 
Moreover, for $\sigma=2$ we expect all scaling laws to agree with short range epidemics. 
The scaling for the latter, corresponding to $\eta=0.58435(50)$, is indicated by the 
dashed-dotted line.
Similar plots were obtained for all $\sigma \in [1.5,2.5]$. Results for 
\be
   \sigma=\sigma_C\equiv 43/24 = 1.7916\ldots
\ee
are e.g. shown in Fig.~\ref{N_crit-sigmaC}.

According to Ref.~\cite{Linder}, the ordinary (short range, GEP) percolation universality class
prevails for all $\sigma >\sigma_C$. The main reason for this is that 
\be
   \eta_{\rm pair} = 2-\sigma
\ee
is supposed to be an exact result \cite{Janssen99} in the intermediate critical phase.
The latter seems indeed supported by Fig.~\ref{pair.fig}, if we interpret the deviations 
seen there as finite time corrections. While this could be consistent with Fig.~\ref{N_crit-sigmaC},
it would be very hard to reconcile this conjecture with Fig.~\ref{N_crit-sigma2}. If there 
is no singularity at $\sigma=2$, it would be very hard to understand the huge finite time
corrections seen in the latter.

\begin{figure}
\includegraphics[width=0.53\textwidth]{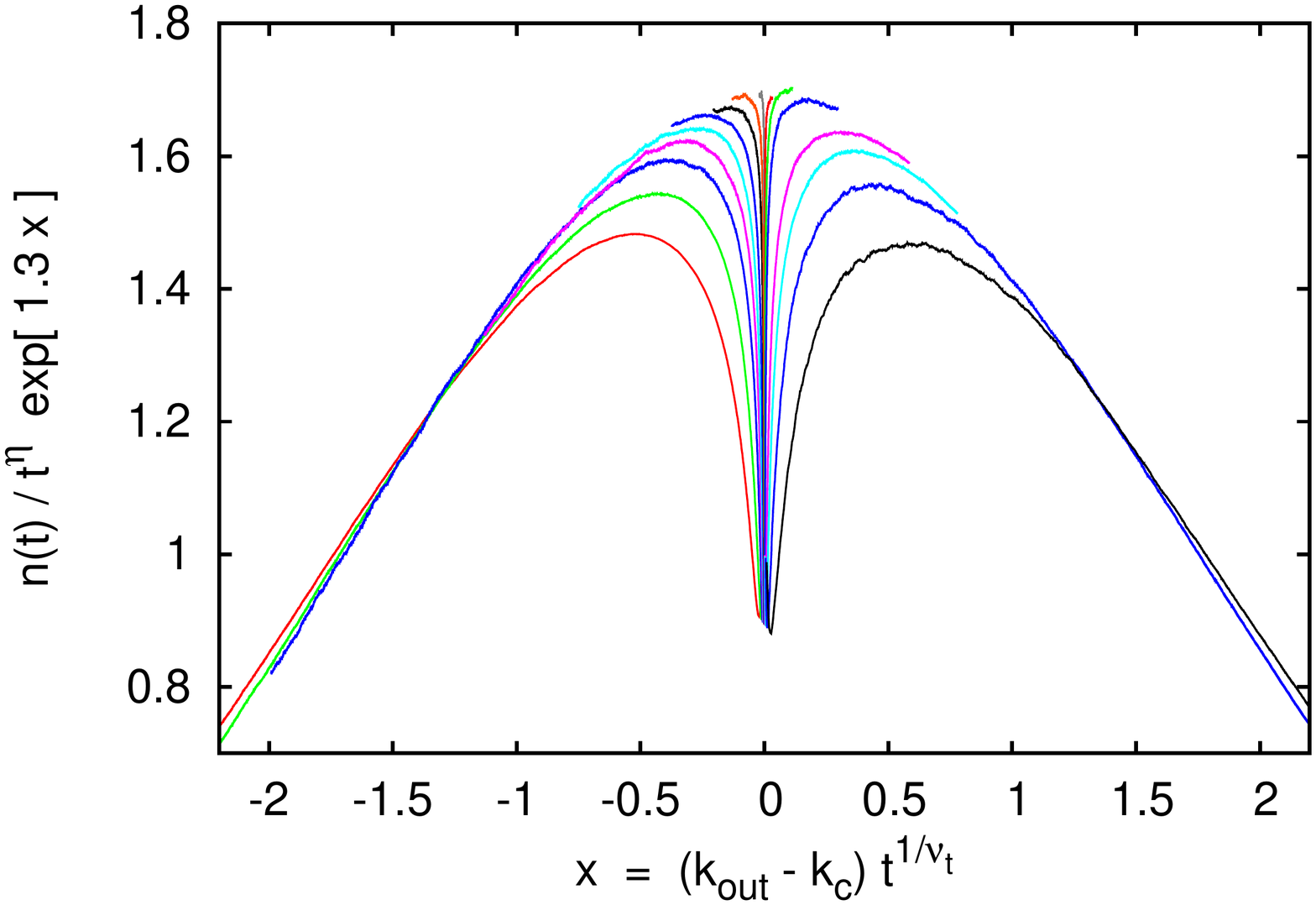}
\includegraphics[width=0.53\textwidth]{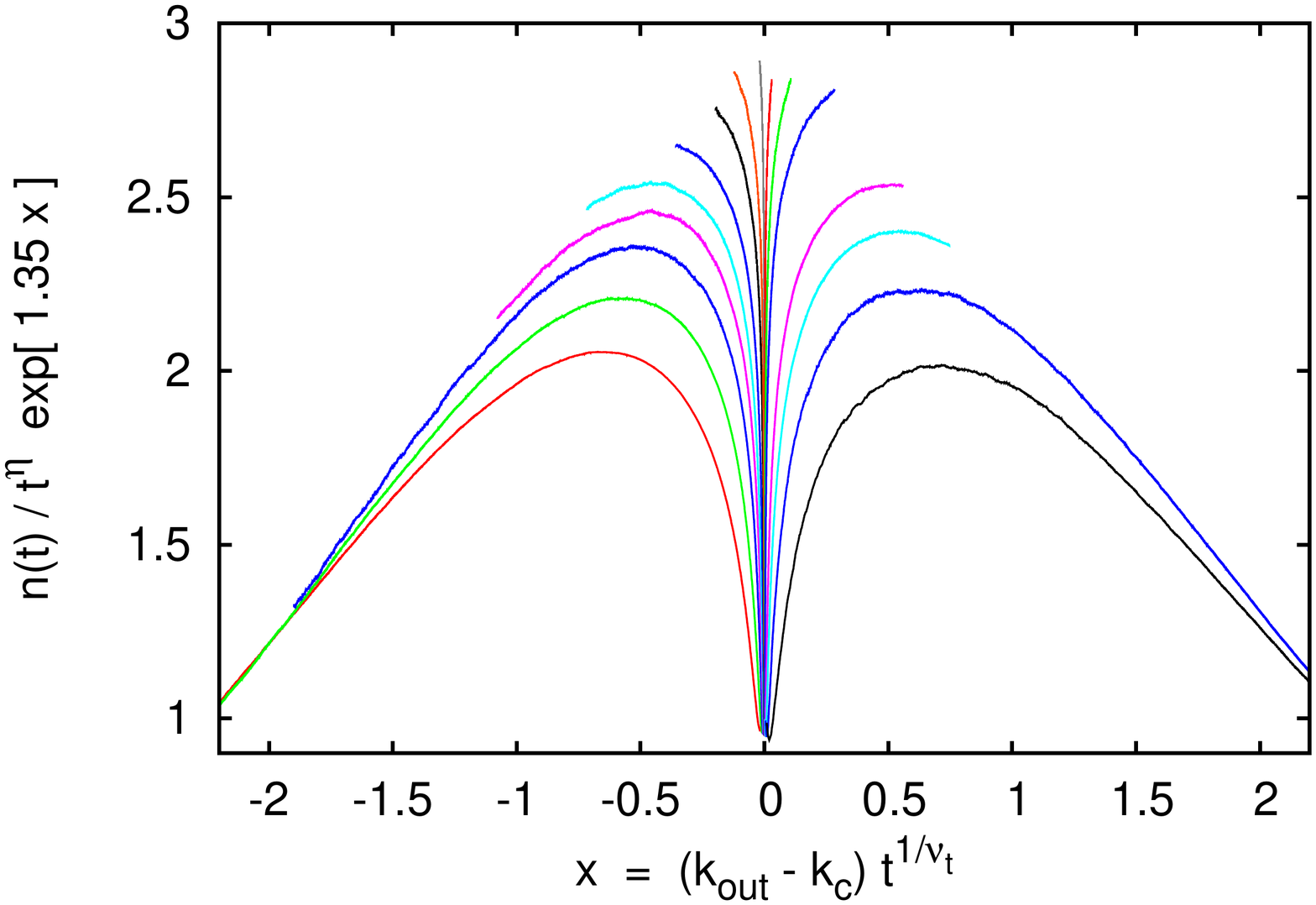}
\caption{(Color online) Collapse plots analogous to Figs.~\ref{collaps-13} and \ref{collaps-bp}.
   In panel a the exponent $\eta = 0.64$ suggested by Fig.~\ref{N_crit-sigma2} is used
   (and $\nu_t$ is fitted as $\nu_t = 1.497$), while in panel b the GEP exponents $\eta = 0.5844$ 
   and $\nu_t = 1.5078$ are used. The curves correspond to the following values of $k_{\rm out}$: 
   1.492, 1.496, 1.498, 1.499, 1.4995, 1.500, 1.5004, 1.50055, 1.50065, 1.5007, 1.5008, 1.5010, 
   1.5020, 1.5030, 1.5050, and 1.510 (from left to right).}
\label{collaps-sigma2}
\end{figure}

A further argument against the conjecture that ordinary GEP universality extends to 
$\sigma<2$ comes from collapse plots similar to Figs.~\ref{collaps-13} and \ref{collaps-bp}.
In Fig.~\ref{collaps-sigma2} we show twice the same data for $\sigma=2$,
once using the critical exponent $\eta = 0.64$ suggested by Fig.~\ref{N_crit-sigma2} (panel a), 
and then using the GEP exponents $\eta = 0.5844$ and $\nu_t = 1.5078$ (panel b). At first sight,
it seems that panel a shows the better collapse. But we should not forget that a good collapse
is only demanded for large $t$, i.e. in the wings of the figure ($|x| >1$, say). There, 
panel b shows the better collapse. But the scaling function $F(x)$ -- the envelope 
of the individual curves -- has a parabolic shape only in panel a. In panel b it seems to
be tent-like, or has an even worse cusp singularity at $x=0$.

We conclude thus that the universality class does indeed change at $\sigma = 2$. The 
critical exponents there are still those of the GEP (as suggested also by Fig.~\ref{Psurv}),
but the scaling functions become singular, presumably due to logarithmic corrections to 
scaling. 

Very puzzling is that some exponents ($\eta_{\rm pair}, \nu_t,\beta$) seem to 
keep their GEP values down to $\sigma_C$, while others ($\eta,z,\nu$) seem to change at $\sigma=2$.
For $\delta$, the situation is unclear.

It is not entirely clear why none of these problems were seen in the simulations of \cite{Linder}.
It is true that these authors used much smaller statistics (5000 runs for each parameter set,
while each curve in the figures shown above is based on $10^5$ to $10^8$ runs). They also 
used much smaller lattices, for which finite lattice effects became a problem for large $t$,
while the above results are free of any finite lattice corrections. Still, the effects seen
e.g. in Figs.~\ref{N_crit-sigma2} and \ref{N_crit-sigmaC} are hard to miss.

\subsection{Finite lattice simulations at $\sigma=\sigma_C$}

Finally, we performed also simulations on finite lattices, in order obtain independent 
estimates for the fractal dimension of the percolation cluster and for exponents $\beta$ 
and $\nu$. In I we had refrained from such simulations, because it was not clear how 
finite size effects should be described in the Kosterlitz-Thouless type transition that 
holds there at $\sigma=d$. For $d=2$, however, the transition is a standard second order 
phase transition (except for possible logarithmic corrections), and we can assume that the 
usual finite size scaling (FSS) applies.

In the following we show only results for $\sigma=\sigma_C$, since there the discrepancy
between the scenarios proposed in \cite{Linder} and in the present paper is most clear. 
At face value, the simulations presented in the last subsection suggest that $\beta >5/36$
and $D_f < 91/48$ ($D_f$ is the fractal dimension of the cluster of ``removed" sites), 
while according to \cite{Linder} these deviations should vanish
in the scaling limit $L,t\to\infty$. In the above simulations we studied the limit
$L\to\infty$ for finite $t$, while simulations on finite lattices allow us to study the 
limit $t\to\infty$ for finite $L$. It is hoped that both limits together can clarify 
the situation better than either limit by itself.

\begin{figure}
\includegraphics[width=0.65\textwidth]{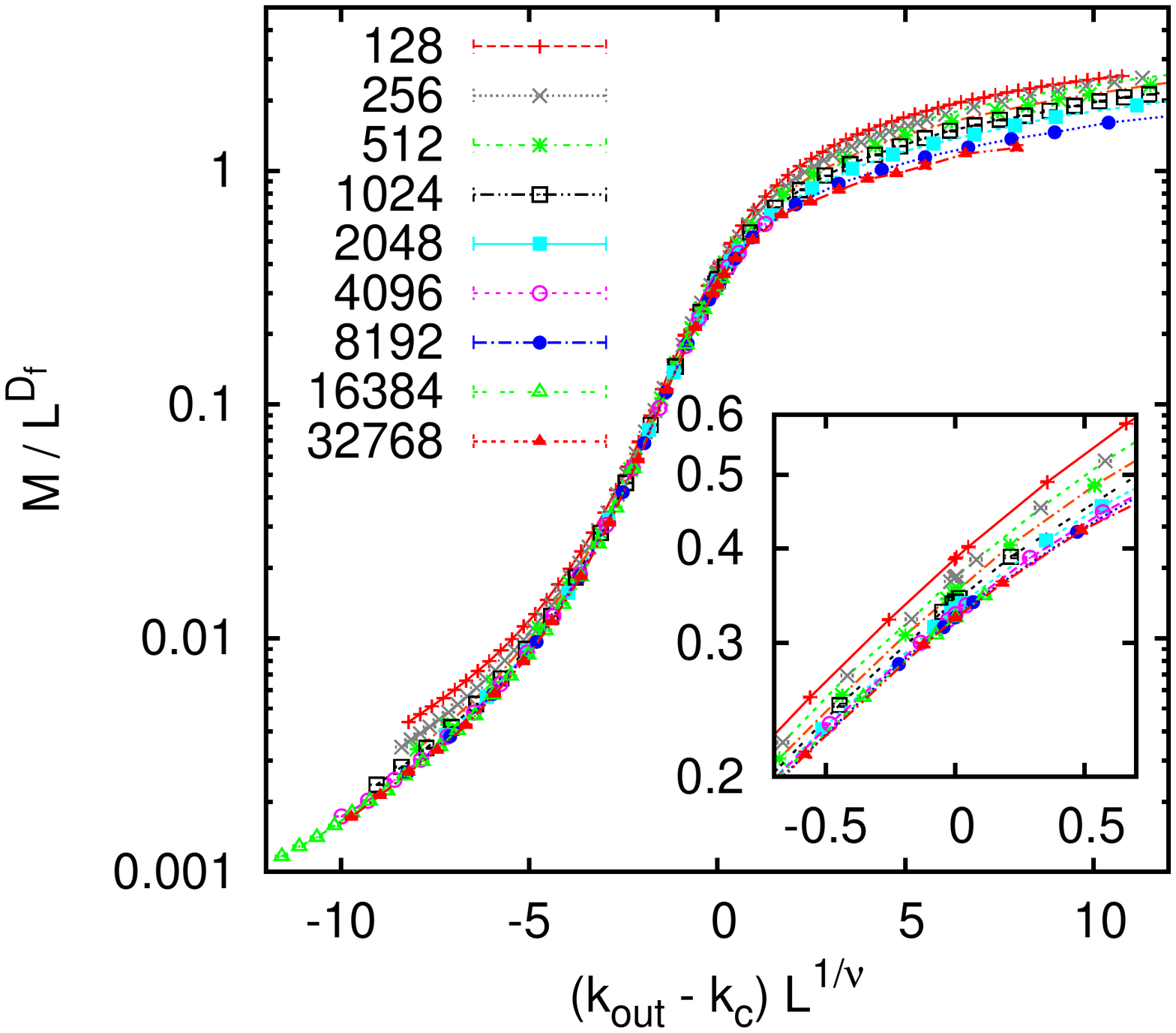}
\caption{(Color online) Collapse plot for the cluster masses $M$ on lattices of size $128\times 128$
   to $32768\times 32768$. The values of $D_f$ and $\nu$ used for this plot are 1.71 and 1.42, 
   both of which are significantly different from the values for ordinary percolation.
   The insert shows that the systematic deviations from scaling in the central region of 
   the plot are mainly for small $L$, indicating that the data do show asymptotic FSS with 
   these exponents.}
\label{fss-collaps}
\end{figure}

In Fig.~\ref{fss-collaps} we show a data collapse based on the FSS ansatz
\be
   M(L,k_{\rm out}) = L^{D_f} G[(k_{\rm out}-k_c) L^{1/\nu}],
\ee
where $M$ is the mass of the cluster of ``removed" (i.e. previously infected) sites. 
We see substantial scaling violations far away from the critical point, which was of 
course to be expected. The main plot of the figure suggests that these scaling 
violations are very small in the central (scaling) region of the plot. The insert
shows that this is not quite true, but that the main violations come from small
values of $L$. The data for large $L$ suggest that 
\be
     D_f = 1.715(30)\;,\qquad \nu = 1.42(3).
\ee
From these we obtain $\beta = (2-D_f)\nu/2 = 0.202(20)$.
These values are significantly different from their values for ordinary percolation,
but are fully consistent with the values obtained in the last subsection.

\begin{figure}
\includegraphics[width=0.53\textwidth]{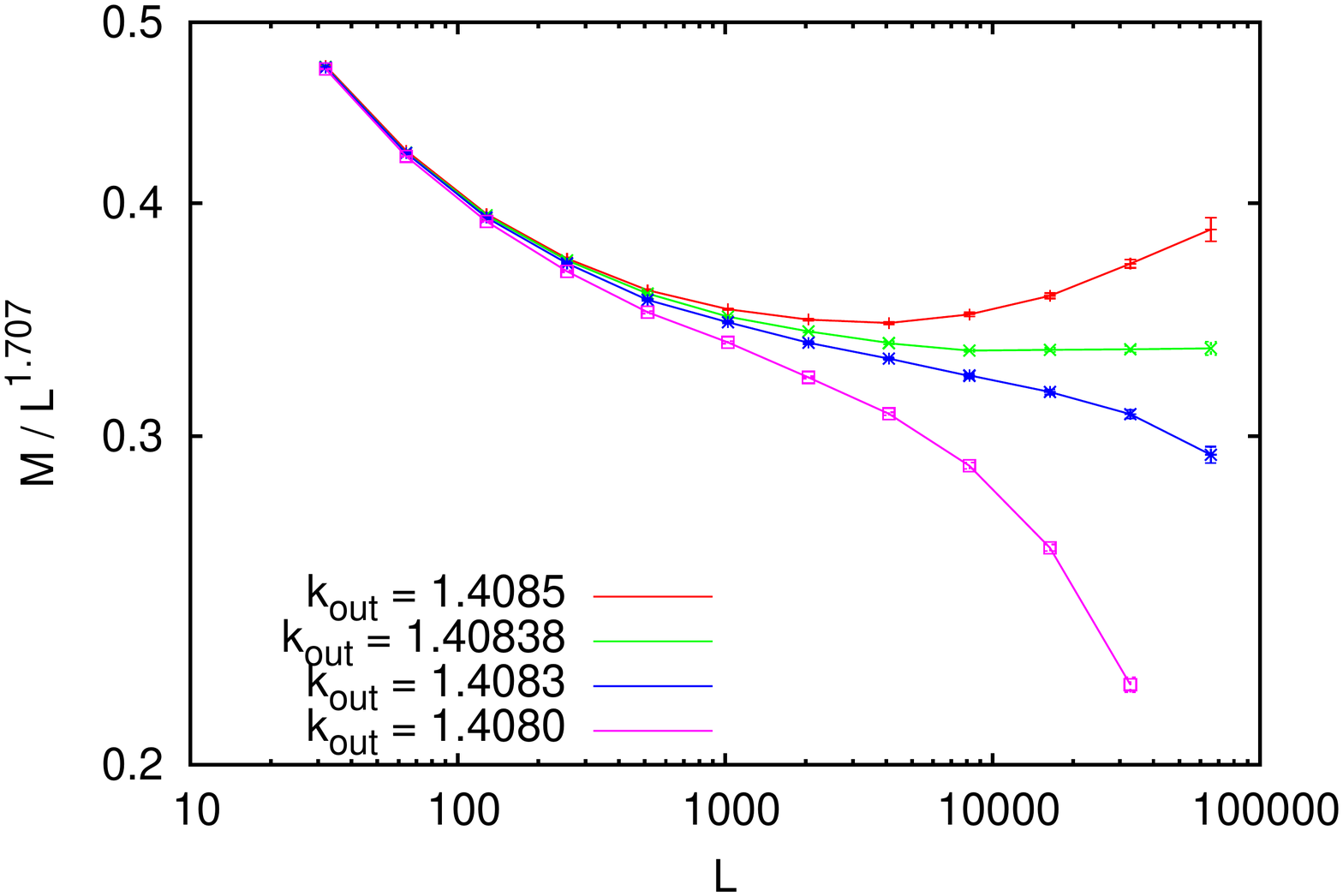}
\caption{(Color online) Log-log plots of $M(L,k_{\rm out})/L^{1.707}$ against $L$,
   for four values of $k_{\rm out}$. If the exponents are properly chosen, the curve for 
   $k_{\rm out}=k_C$ should become horizontal for $L\to\infty$.}
\label{D_f}
\end{figure}

While this is a further indication that the cross-over to ordinary percolation does not 
happen at $\sigma_C$, it is not very convincing since scaling corrections are notoriously
easy to miss in such data collapse plots. Therefore we also show in Fig.~\ref{D_f} log-log
plots of $L^{-D_f}M(L,k_{\rm out})$ against $L$, for several values of $k_{\rm out}$ 
close to $k_C$. Asymptotically, these curves should veer up (down), if $k_{\rm out}$ is 
larger (smaller) than $k_C$. From Fig.~\ref{N_crit-sigmaC} (and from the corresponding
plot for $P_{\rm surv}$ we see two main alternatives.
\begin{itemize}
\item  Either (a): $\nu_t$ assumes the GEP value, in which case $k_C \leq 1.40835$;

\item Or (b): $\nu_t \approx 0.67$, in which case $k_C = 1.408380(15)$.

\end{itemize}

The data in Fig.~\ref{N_crit-sigmaC} clearly indicate that the value 1.4085 is supercritical,
while 1.408 and 1.4083 are subcritical, in perfect agreement with Fig.~\ref{N_crit-sigmaC}.
But they also suggest strongly that $k_{\rm out}=1.40838$ is very close to critical.
In this case the fractal dimension would be $D_f = 1.707(7)$, which is definitely smaller 
than the value $91/48 = 1.896$ of ordinary percolation.
On the basis of Fig.~\ref{N_crit-sigmaC} alone, a critical value 1.40835 seems unlikely but not excluded. 
But it would give an even smaller value of $D_f$ and is thereby clearly excluded. Our final 
estimate is $D_f = 1.707(17)$, where the uncertainty is mainly due to the uncertainty of $k_c$.

\section{Discussion and Conclusions}

This work was triggered by recent discussions in the physics literature of spatially embedded 
networks with long range connections \cite{Barthelemy,Sen01,Moukarzel,Sen02,Goswami11,Kosmidis,Daqing,Emmerich}
and supercritical epidemic processes leading to such networks \cite{Castillo,Mancinelli,Brockmann}.
Our simulations showed that most of these speculations are obsolete, but in retrospect this 
was to be expected. The correct topology of such networks was known {\it rigorously} since 2004
\cite{Biskup-2004}. If the probability to infect a neighbor at distance $\bf x$ decays as 
$|\bf x|^{-s-\sigma}$ with $0<\sigma < s$, then the number of nodes reached by a path of 
length $t$ increased like a stretched exponential with known (and non-trivial) exponent.
Our simulations fully agree with this prediction, in spite of the notorious difficulty
to fit stretched exponentials.

Another class of problems is concerned with the supercritical behavior at $\sigma=d$. 
It was observed already in the late 1960's that the case $\sigma=d$ is special. For instance,
the 1-d Ising model shows a phase transition for $\sigma <1$, while it has no transition 
for $\sigma >1$. For 1-d percolation, a seminal early result was that there is a {\it discontinuous}
phase transition at $\sigma=1$ with Kosterlitz-Thouless like behavior in the supercritical
regime \cite{Aizenman}. No analogous result was known in two dimensions, although there 
are conjectures \cite{Benjamini-2001} that there also might exist continuously varying 
exponents in the supercritical phase. In the present paper we verify this conjecture and 
suggest various scaling laws related to it.
 
In contrast to the 1-d case, the percolation transition in the 2-d case with $\sigma=2$
is continuous. Indeed, there are very strong theoretical arguments that the critical 
exponents at this point are exactly those of ordinary percolation or, as far as temporal 
aspects are concerned, of  
the ``general epidemic process" (GEP). On the other hand, there is a long-standing 
debate for the Ising model (where the situation in this respect should be very similar)
whether the ordinary short-range behavior should end at $\sigma=2$ or extend some way
into the region $\sigma<2$ \cite{Fisher72,Sak,Yamazaki,Gusmao,Honkonen,Luijten02}. 
This debate is mostly centered around field theoretic arguments, but the most 
recent simulations \cite{Luijten02} seemed to have settled the problem: The 
ordinary short-rang behavior extends a finite amount into the region $\sigma <2$. 
The same conclusion was reached for percolation in \cite{Linder}. 

Our present simulations show rather convincingly that this is wrong -- at least
for percolation, but the theoretical analogy suggests also for the Ising model.
There is a singularity at $\sigma=2$, and at least {\it some} of the critical
exponents are different for all $\sigma < 2$ from those for ordinary percolation.
While we are rather confident about this basic claim \footnote{Notice that one cannot 
use the huge corrections to scaling seen e.g. in Fig.~\ref{N_crit-sigma2} to argue away
the problem. If there were no singularity at $\sigma=2$, there would be no reason 
for any large corrections to scaling.}, details are much harder to 
pin down due to huge corrections to scaling. This represents the main open problem
related to the present paper.

The field theory that had given rise to the above debate can be treated by 
renormalization group methods near $\sigma=d/3$ \cite{Janssen99}, where a field theoretic 
$\epsilon$-expansion predicts anomalous critical exponents up to first order 
in $\epsilon = \sigma-d/3$. When comparing our simulations with these predictions,
we find again large discrepancies, which might suggest that the $\epsilon$-expansion
is more singular that expected. Again more works would be needed to settle this 
question.

In summary, percolation with long range infection is a fascinating problem. It 
touches basic questions of renormalization group theory, it has applications 
to real-world epidemics, and it sheds light on the structure of real-world 
complex networks. And, finally, it still shows a number of open questions after 
having been studied for more than 40 years.

\section*{Acknowledgments}

For very helpful discussions I want to thank Aicko Schumann and Deepak Dhar.
I also want to thank the latter for the kind hospitality at the Tata Institute
of Fundamental Research in Mumbai, where part of this work was done.
The work was begun at the Complexity Science 
Group at the Universality of Calgary. Finally my thanks go to Haye Hinrichsen and Hans-Karl 
Janssen for illuminating correspondence.

\bibliography{mm}

\end{document}